\documentclass[aps,superscriptaddress,groupedaddress,twocolumn,floatfix,10pt]{revtex4}
\usepackage{natbib}
\usepackage{graphicx} 
\usepackage{dcolumn} 
\usepackage{bm} 
\usepackage{amssymb,amsmath} 
\usepackage{textcomp} 
\usepackage[utf8x]{inputenc}
\usepackage{braket}
\usepackage{amsmath}
\usepackage{placeins}
\usepackage{xcolor}
\usepackage{pdfpages}
\usepackage{soul}
\usepackage{hyperref}       
\usepackage{url}            
\usepackage{booktabs}       
\usepackage{todonotes} 
\usepackage{changebar}
\usepackage{changes}
\definechangesauthor{AB}

\begin{document}

\pagestyle{empty}

\title{Stock market microstructure inference via multi-agent reinforcement learning}

\author{
Johann Lussange\\
Laboratoire des Neurosciences Cognitives, INSERM U960, Département des Études Cognitives,\\
École Normale Supérieure, 29 rue d'Ulm, 75005, Paris, France.\\
Ivan Lazarevich\\
Laboratoire des Neurosciences Cognitives, INSERM U960, Département des Études Cognitives,\\
École Normale Supérieure, 29 rue d'Ulm, 75005, Paris, France.\\
Lobachevsky State University of Nizhny Novgorod, \\
23 Gagarina av., 603950, Niznhy Novgorod, Russia. \\
Sacha Bourgeois-Gironde\\
Institut Jean-Nicod, UMR 8129, Département des Études Cognitives,\\
École Normale Supérieure, 29 rue d'Ulm, 75005, Paris, France.\\ 
Laboratoire d'Économie Mathématique et de Microéconomie Appliquée, EA 4442,\\
Université Paris II Panthéon-Assas, 4 rue Blaise Desgoffe, 75006, Paris, France.\\
Stefano Palminteri\\
Laboratoire des Neurosciences Cognitives, INSERM U960, Département des Études Cognitives,\\
École Normale Supérieure, 29 rue d'Ulm, 75005, Paris, France.\\
Boris Gutkin\\
Laboratoire des Neurosciences Cognitives, INSERM U960, Département des Études Cognitives,\\
École Normale Supérieure, 29 rue d'Ulm, 75005, Paris, France.\\ 
Center for Cognition and Decision Making, Department of Psychology,\\
NU University Higher School of Economics, 8 Myasnitskaya st., 101000, Moscow, Russia.\\
}

\date{\today}

\begin{abstract} 

Quantitative finance has had a long tradition of a bottom-up approach to complex systems inference via multi-agent systems (MAS). These statistical tools are based on modelling agents trading via a centralised order book, in order to emulate complex and diverse market phenomena. These past financial models have all relied on so-called zero-intelligence agents, so that the crucial issues of agent information and learning, central to price formation and hence to all market activity, could not be properly assessed. In order to address this, we designed a next-generation MAS stock market simulator, in which each agent learns to trade autonomously via model-free reinforcement learning. We calibrate the model to real market data from the London Stock Exchange over the years $2007$ to $2018$, and show that it can faithfully reproduce key market microstructure metrics, such as various price autocorrelation scalars over multiple time intervals. Agent learning thus enables model emulation of the microstructure with greater realism. 

\end{abstract}

\maketitle

\section{Background}

\textit{Past research}: The field of research in finance and economics has historically explored various types of quantitative models for its statistical inference of stock market data. Among these, we can briefly mention two general classes of models. The first and most encountered ones are autoregressive time-series models aimed at prediction of future values from past history~\cite{Greene2017}. The second one, MAS~\citep{Wellman2017} (agent-based models and related methods such as order book models~\citep{Huang2015,Biondo2019,Sirignano2019}, and dynamic stochastic general equilibrium models~\citep{Sbordone2010}), rather pertain to the causal sources of financial markets activity. The latter may be applied to both high and low-frequency trading~\citep{Wah2013,Aloud2014} and to the study of supply and demand~\citep{Benzaquen2018} in the form of game theory~\citep{ErevRoth2014} and the so-called minority game~\citep{Demartino2006}. From a regulatory point of view, it has an ever-increasing role to play~\citep{Boero2015}, in particular wrt. macroeconomics~\citep{Gualdi2015}. 

\textit{New trends}: But recent trends have potentially given MAS research in finance a whole new range of realism, which to our knowledge is yet unexplored. These trends emerge from the association of two present-day major scientific breakthroughs: i- the steady advances of cognitive neuroscience and neuroeconomics~\citep{Eickhoff2018,Konovalov2016}, and ii- the progress of reinforcement learning due to a general renewed attention on machine learning methods and especially multi-agent learning~\citep{Silver2018,Silver2017}. This has been accompanied on both ends with the emergence of reinforcement learning algorithms incorporating decision-theoretic features from neuroeconomics~\citep{Lefebvre2017,Palminteri2015}, and neuroscience models approached from the angle of reinforcement learning~\citep{Duncan2018,Momennejad2017}. These developments offer a way to go beyond the former generation of MAS with zero-intelligence agents~\citep{Gode1993}, and their potential financial applications~\citep{Hu2019,Neuneier1997,Deng2017} have very recently started to be extended to the class of order book models, coupled with reinforcement learning~\citep{Spooner2018}. 

\textit{Our contribution}: In order to address this, we have developed a next generation stock market simulator based on a MAS architecture, where each agent represents an economic investor trading via a centralised order book. In such a model, the simulated agents have three novel features: i- each agent learns to both forecast and trade by independent reinforcement learning algorithms in a fully autonomous way; ii- each agent learns a pricing strategy for this forecasting and trading that is more or less chartist (i.e relying on market price) or fundamental (i.e relying on intrinsic economical value); iii- each agent can be endowed with certain traits of behaviour, cognition, and learning pertaining to behavioural economics, thanks to the reinforcement learning framework and its direct correspondance with decision theory. These features provide a whole new level of realism in simulated data and its emulation of real stock markets data.

\section{Stylised facts}

Over time, quantitative research has discovered certain patterns in financial data, which seem to constrain the domain of the efficient market hypothesis and indirectly posit some sort of market memory. These were called \textit{stylised facts} and gradually discovered over the nineties: Kim-Markowitz~\cite{Kim1989}, Levy-Levy-Solomon~\cite{Levy1996,Levy1994,Levy1995,Levy1996b,Levy1996c,Levy1997,Levy2000}, Cont-Bouchaud~\cite{Cont2000}, Solomon-Weisbuch~\cite{Solomon2000}, Lux-Marchesi~\cite{Lux1999,Lux2000}, Donangelo-Sneppen~\cite{Donangelo2000,Donangelo2000b,Bak1999,Bak2001}, Solomon-Levy-Huang~\cite{Huang2000}. Their emulation in MAS financial research has since then been an active topic of research~\cite{Lipski2013,Barde2015,Bouchaud2018}. What is remarkable is that these stylised facts can generalise to cross-asset markets, and show remarkable time invariance. Understanding such universal market features also pertains to the exogenous or endogenous causes to price formation~\cite{Dodonova2018,Naik2018}. Implicit consequences of these stylised facts have fed numerous discussions pertaining to the validity of market memory~\cite{Cont2005,Cristelli2014} and the extension of the efficient market hypothesis~\cite{Fama1970,Bera2015}. Their definite characteristics has varied ever so slightly over the years and across literature, but the most widespread and unanimously accepted stylised facts can in fact be grouped in three broad, mutually overlapping categories that we here sum up:

\vspace{1mm}

i- \textit{Non-gaussian returns}: the returns distribution is non-gaussian and hence asset prices should not be modelled as brownian random walks~\cite{Potters2001,Plerou1999}, despite what is taught in most text books, and often applied in sell-side finance. In particular the real distributions of returns are dissimilar to normal distributions in that they are: i- having fatter tails and hence more extreme events, with the tails of the cumulative distribution being well approximated~\cite{Cristelli2014,Potters2001} by a power law of exponent belonging to the interval $[2,4]$ (albeit this is the subject of a discussion~\cite{Weron2001,Eisler2006} famously started by Mandelbrot~\cite{Mandelbrot1963} and his Levy stable model for financial returns), ii- negatively skewed and asymmetric in many observed markets~\cite{Cont2001}, with more large negative returns than large positive returns, iii- platykurtic and as a consequence having less mean-centered events~\cite{Bouchaud1997}, iv- with multifractal $k$-moments, so that their exponent is not linear with $k$, as seen in~\cite{Ding1993,Lobato1998,Vandewalle1997,Mandelbrot1997}. 

\vspace{1mm}

ii- \textit{Clustered volatilities}: market volatility tends to aggregate or form clusters~\cite{Engle1982}. Therefore compared to average, the probability to have a large volatility in the near-future is greater if it was large also in the near-past~\cite{Lipski2013,Devries1994,Pagan1996}. Regardless of whether the next return is positive or negative, one can thus say that large (resp. small) return jumps are likely followed by the same~\cite{Mandelbrot1963}, and thus display some sort of long memory behaviour~\cite{Cont2005}. Because volatilities and trading volumes are often correlated, we also observe a related volume clustering. Indirectly, this has long-range implications on the dynamics of meta-orders, and comprises the \textit{square-root impact law}~\cite{Bouchaud2018} (growth in square-root of orders impact with traded volumes). 

\vspace{1mm}

iii- \textit{Decaying auto-correlations}: the auto-correlation function of the price returns of financial time series are basically zero for any value of the auto-correlation lag, except for very short lags (e.g. half-hour lags for intraday data) because of a mean-reverting microstructure mechanism for which there is a negative auto-correlation~\cite{Cont2001,Cont2005}. This is sometimes feeding the general argument of the well-known Efficient Market Hypothesis~\cite{Fama1970,Bera2015} that markets have no memory and hence that one cannot predict future prices based on past prices or information~\cite{Devries1994,Pagan1996}. According to this view, there is hence no opportunity for arbitrage within a financial market~\cite{Cristelli2014}. It has been observed however that certain non-linear functions of returns such as squared returns or absolute returns display certain steady auto-correlations over longer lags~\cite{Cont2005}. 

\vspace{3mm}

\section{Reinforcement learning}

\subsection{Basic concepts}

\textit{Three variables}: We here outline a brief overview of reinforcement learning theory. Unlike supervised or unsupervised learning, reinforcement learning is another paradigm of machine learning based on the definition of reward or reinforcement. For a thorough study of the subject, we refer the reader to~\cite{SuttonBarto,Wiering2012,Csaba2010}. Three main variables must first and foremost be specified for any reinforcement learning agent: the possible states $s \in \mathcal S$ of the environment in which the agent evolves and over which it has no control, the possible actions $a \in \mathcal A$ that the agent can effectively control and perform in its environment, and the reward or reinforcement $r \in \mathbb R$ proper to that agent. Note that like most other machine learning approaches, reinforcement learning assumes Markov state signals, a state signal that succeeds in retaining all relevant information being said to be Markov, or to have the Markov property if and only if $\forall s', r'$ and histories $s_t, a_t, r_t, s_{t-1}, a_{t-1}, r_{t-1}, \cdots, s_1, a_1, r_1, s_0, a_0$, we have:
\begin{eqnarray}
&& Pr \{ s_{t+1} = s', r_{t+1} = r' | s_t, a_t \}  \nonumber \\ 
&=& Pr \{ s_{t+1} = s', r_{t+1} = r' | s_t, a_t, r_t, \cdots,  r_1, s_0, a_0 \}. \nonumber
\end{eqnarray}

\vspace{1mm}

\textit{Three functions}: Maximising its total reward over time is the ultimate goal of the agent. In order to do so, it needs to learn an optimal behaviour in that environment, i.e. what are the best actions $a$ to perform for each given state $s$. For this, a function called the \textit{policy}:
\begin{eqnarray}
\pi(s,a) = Pr (a | s)
\end{eqnarray}
\noindent is initialised with equiprobable actions in the beginning, but updated via exploration by the agent to find an optimal policy, denoted $\pi^{\ast}(s,a)$. Hence at time $t$, the agent tries a new action $a_t$ in a state $s_t$ and then observes its associated reward, so as to update the probabilities of $\pi(s,a)$ accordingly. In order to gauge and assess the amplitude of this reward, the agent may work with the so-called \textit{state-value function}: 
\begin{eqnarray}
V(s)= \mathbb E [ \sum_{k=0}^{\infty} \gamma^k r_{t+k+1} | s_t=s ]
\end{eqnarray}
\noindent which is linked with two functions called the transition probability $\mathcal P_{ss'}^a = Pr \{ s_{t+1} = s' | s_t=s, a_t=a \}$ and the expected value $\mathcal R_{ss'}^a = \mathbb E [ r_{t+1} | s_t=s, a_t=a, s_{t+1}=s' ]$, where $0 < \gamma < 1$ is a discount parameter related to the concept of delayed reward. Or the agent may alternatively work with the so-called \textit{action-value function}: 
\begin{eqnarray}
Q(s,a)= \mathbb E [ \sum_{k=0}^{\infty} \gamma^k r_{t+k+1} | s_t=s, a_t=a ]. 
\end{eqnarray}

\textit{Three families}: The way to learn the optimal policy $\pi^{\ast}(s,a)$ practically in reinforcement learning is usually separated in several families of algorithms. In the so-called \textit{model-based} methods (such as \textit{Dynamic Programming}), the agent may after each performed action $a$ keep record of its associated transition probability $\mathcal P$ and expected value $\mathcal R$, then compute its state-value function $V$, in order to update its policy $\pi$. In \textit{model-free} methods (such as \textit{Monte Carlo} and \textit{Temporal Difference}, which can be both united via eligibility traces in the form of TD($\lambda$)), the agent may simply after each performed action $a$ keep record of its action-value function $Q$ in order to update its policy $\pi$. Thus model-based simply means that the agent keeps a record of the rewards associated with a model it updates of its environnement, while in model-free reinforcement learning, the agent doesn't keep such a model. Finally, \textit{direct policy search} methods are sometimes used when the agent bypasses the record of state or action-value functions, and updates the probabilities of state-action pairs of the policy after receiving the associated reward directly. Searching the policy space can be done via \textit{gradient-based} and \textit{gradient-free} methods. 

\vspace{1mm}

\textit{Three features}: Like many other machine learning methods, reinforcement learning draws its inspiration from biology: optimal behaviour is thus not learned directly as such but rather a reward (or reinforcement) is predefined and the behaviour is indirectly learned through trial and error, formally defined as an \textit{exploration vs. exploitation} process. For the agent faces a dilemma, namely whether it should exploit a given action it knows yields a good reward, or explore more its environment and action set so as to find even better rewards. Certain methods have been proposed to answer this exploration versus exploitation dilemma, among which the $\epsilon$-greedy method (a small probability $\epsilon$ is chosen to explore a random action at each time step while exploiting the best action otherwise; when $\epsilon=0$ this algorithm is simply called the greedy method), softmax method (the action to explore is not purely random but graded from known best to worse according to a temperature parameter), or pursuit method (continually pursuing the action that is greedy) are most often encountered. Furthermore a series of actions is sometimes necessary over prolonged periods of time before reaching the reward, and reinforcement learning thus deals with the concept of \textit{delayed reward} or temporal credit assignment, formally defined as the sum of discounted rewards over time $\sum_{k=0}^{\infty} \gamma^k r_{t+k+1}$, which we already saw in the previous state-value $V(s)$ and action-value $Q(s, a)$ functions. Finally, a last central feature proper to reinforcement learning is the so-called \textit{curse of dimensionality}, namely that the number of state-action pairs that the agent must explore becomes quickly computationally intractable in practical applications. These aspects of exploration versus exploitation, delayed reward, and dimensionality are central features of reinforcement learning, and active domains of research. 

\subsection{Fields of research}

As we just saw, most reinforcement learning research has clustered around these three features of reinforcement learning mentioned above, often with some overlap:

\vspace{2mm}

i- \textit{Exploration vs. exploitation}: This dilemma has been addressed by different policy learning methods, such as \textit{policy gradient methods}~\cite{Sutton2000,Silver2014} seeking to optimize the control policy with respect to the return by gradient descent, or \textit{actor-critic methods}~\cite{Grondman2012}, where an actor controls the behavior of the agent (policy-based reinforcement learning) and a critic evaluates the action taken by the agent (value-based reinforcement learning). Another more recent approach is \textit{meta-reinforcement learning}~\cite{Wang2018}, which deals with ``learning how to learn" in order to improve a faster generalization, especially at different time-scales~\cite{Duan2016} (see also \textit{zero-shot} or \textit{few-shots reinforcement learning}).Together with this approach, we should mention \textit{transfer reinforcement learning}~\cite{Wiering2012}, which is to transfer the experience gathered on one task to another, and \textit{imitation} or \textit{apprenticeship reinforcement learning}, which is to learn a task from observation of another agent. We can also mention the increasing role played by \textit{multi-agent learning} and \textit{self-play reinforcement learning}~\cite{Heinrich2017}, which deals with learning a policy by playing against another agent that also learns. \textit{Multitask reinforcement learning} seeks to learn many tasks and exploit their similarities in order to improve single-task performance. It is related to the now famous \textit{asynchronous reinforcement learning}~\cite{Mnih2016}, which executes in parallel many instances of an agent while using a shared model in order to obtain data diversification, and \textit{modular reinforcement learning}~\cite{Andreas2017}, which learns the policy of a task by dividing it into smaller subtasks and reconstructing their individual policies. Another well-known approach is \textit{Monte Carlo Tree Search reinforcement learning}~\cite{Silver2016}, which determines the best action via a tree search relying on random sampling of the search space. We can also mention the important fields of \textit{lifelong reinforcement learning}~\cite{Tessler2016}, which deals with learning a large amount of sequential tasks, and \textit{hierarchical reinforcement learning}~\cite{Bhatnagara2006} which regroups the agent actions in more general tasks. Finally, we can name \textit{hybrid reinforcement learning}~\cite{Liang2017} (or \textit{human-in-the-loop reinforcement learning}), which deals with human interference in the algorithmic learning process in order to improve it (c.f. intelligent driving). 

ii- \textit{Temporal credit assignment}: This is an issue that deals with the general specification and definition of the agent's reward and return. Apart from the aforementioned hierarchical reinforcement learning approach, ongoing research to address this difficulty includes \textit{shaping rewards}~\cite{Ng1999}, which deals with incorporating background knowledge on sub-rewards in order to improve convergence rates, \textit{inverse reinforcement learning}~\cite{Abbeel2010}, which seeks to extract the reward function out of the observation of the (optimal) behavior of another agent. We can also mention \textit{homeostatic reinforcement learning}~\cite{Keramati2014,Keramati2011}, which defines the reward via a manifold of many sub-rewards, and a state-dependent approach to the definition of the agent's reward~\cite{Bavard2018}. 

iii- \textit{Curse of dimensionality}: This issue arises when dealing with the issue of large-scale MDPs, so that the exploration and hence convergence to an optimal policy $\pi^{\ast}(s,a)$ becomes quickly intractable. In particular, the so-called $Q$-\textit{learning} algorithm~\cite{Watkins1992}, as part of the more general model-free temporal difference or TD-learning was a breakthrough for reinforcement learning, because it reduced drastically the number of states to explore: only the pairs $(s,a)$ need be explored. This curse of dimensionality points to the more general problem of function approximation, which recently led to the interface of reinforcement learning with artificial neural networks in the form of the now famous \textit{end-to-end} or \textit{deep reinforcement learning}~\cite{Silver2016}. Related to this is the ongoing work on \textit{partially observable MDP models}~\cite{Ross2011,Katt2017}, and \textit{adversarial reinforcement learning}~\cite{Pinto2017}, which deals with modeling the noise and uncertainties in state representation via an adversarial agent applying certain specific perturbations to the system.

\vspace{3mm}

\section{Model}

\subsection{General architecture}

The main simulation parameters are the number of agents $I$, the number of traded stocks $J$, the number of simulation time steps $T$ (we consider a time step to correspond to a trading day, a year to correspond to $T_{y}=252$ trading days, a month to $T_{m}=21$ trading days, and a week to $T_{w}=5$ trading days). Usually, we consider statistical features emerging from a number $S$ of simulations. We also model the friction costs via broker fees applied to each transaction set at $b=0.01 \%$, an annual risk-free rate of $R=1 \%$, and an annual stock dividend yield of $D=2 \%$ according to~\cite{DividendYield}. The simulation then follows at each time step $t$ the four general steps described below:

\vspace{2mm}

i- \textit{Initialisation of agents parameters}: The simulation initialises $I$ agents and all their individual parameters at $t=0$. Each agent represents an individual or corporate investor or trader managing its own portfolio over time $t$ via interaction with the market. This portfolio is made of specific stock holdings (equity) and risk free assets (bonds). These agent parameters are described in further detail in section B below. 

\begin{figure}[!htbp]
\begin{centering}
\includegraphics[scale=0.4]{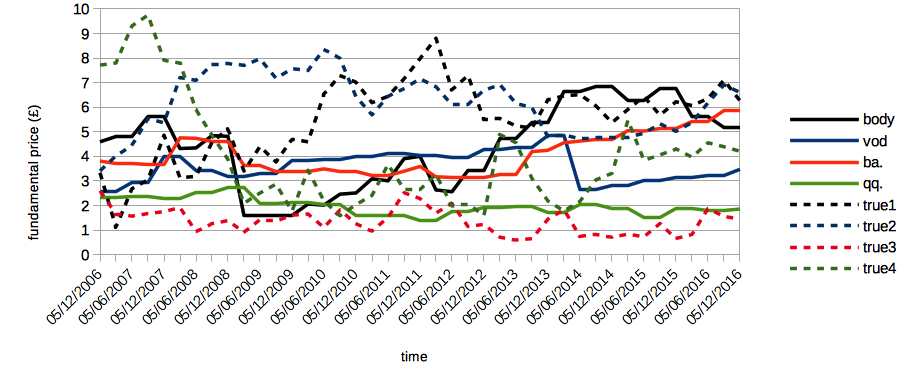}
\caption{\label{Z1} Examples of fundamental values modelled by $\mathcal{T}^{j}(t)$ (black dashed curve) and some agent's approximation thereof as $\mathcal{B}^{i,j}(t)$ (continuous blue, red, and green curves) over a simulated time of one year.}
\end{centering}
\end{figure}

\begin{figure}[!htbp]
\begin{centering}
\includegraphics[scale=0.5]{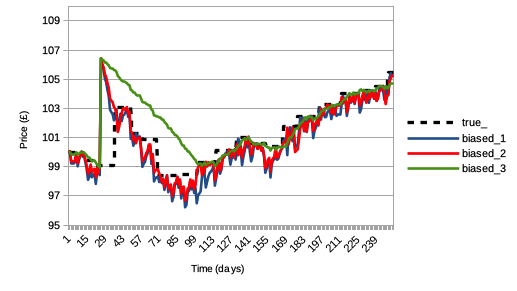}
\caption{\label{Z2} Examples of fundamental values in the London Stock Exchange (cf. symbols as legend) over the years $2006$ to $2016$ represented by enterprise value divided by the number of stock outstanding (continuous curves), and those generated as $J=4$ unscaled time series $\mathcal{T}^{j}(t)$ (dashed curves) by the simulator at time $t=0$.}
\end{centering}
\end{figure}

\vspace{2mm}

ii- \textit{Initialisation of fundamentals}: As in other models~\cite{Franke2011,Chiarella2007}, the simulation initialises all market prices at $P^{j}(t=0)=\$ 100$, and generates $J$ time series $\mathcal{T}^{j}(t)$ as jump processes, which correspond to the prices associated with the fundamental values of the stocks. We model the topology of these fundamental values out of a metric that is often seen in corporate finance and company quarterly reports, called the \textit{enterprise value} (which represents the approximative value at which the company would be bought in mergers and acquisitions for instance), and which we divide by the total number of stock outstanding~\cite{Vernimmen}. These are not fully known by the $I$ agents. Instead, each agent $i$ approximates the values $\mathcal{T}^j(t)$ of stock $j$ according to a proprietary rule~\citep{Murray1994} of cointegration $\mathcal{\kappa}^{i,j} [ \mathcal{T}^{j}(t) ]=\mathcal{B}^{i,j}(t)$. The time series $\mathcal{B}^{i,j}(t)$ are hence the fundamental values of stock $j$ over time $t$ according to agent $i$. Together with such modelled fundamental values $ \mathcal{T}^{j}(t)$, we show on Fig. \ref{Z1} examples of such enterprise values divided by the number of stock outstanding, for various companies traded on the London Stock Exchanges, over the years $2006-2016$. On Fig. \ref{Z2}, we plotted a sample of such modelled fundamental values $ \mathcal{T}^{j}(t)$, together with their approximation $\mathcal{B}^{i,j}(t)$ by some agents. The average annual number of jumps of $\mathcal{T}^j(t)$ is $12.70 \pm 1.85$, the average amplitude $(\mathcal{T}^j(t)-\mathcal{T}^j(t-1))/\mathcal{T}^j(t)$ of these jumps is $5.90 \pm 1.84 \%$, and the average difference between biased and true values $(\mathcal{T}^j(t)-\mathcal{B}^{i,j}(t))/\mathcal{T}^j(t)$ is $2.37 \pm 1.36 \%$ (where $\pm$ terms refer to standard deviations). Each agent thus relies on these two sources of information for its stock pricing strategy: one that is chartist, and one that is fundamental. 

iii- \textit{Agents forecasting and trading}: Each agent autonomously uses two distinct reinforcement learning algorithms to interact with the market, each algorithm being described in further detail in section C below. A first algorithm $\mathcal{F}^{i}$ learns the optimal econometric prediction function for the agent's investment horizon, depending on specific local characteristics of the market microstructure and the agent's fundamental valuation $\mathcal{B}^{i,j}(t)$. It thus outputs this price forecast, which will in turn enter as input the second reinforcement learning algorithm $\mathcal{T}^{i}$. This second algorithm is in charge of sending an optimal limit trading order to a double auction order book~\cite{Mota2016} at this same time step, based on this prediction and a few other market microstructure and agent portfolio indicators. Notably, the transaction order output by $\mathcal{T}^{i}$ is filtered by a function $\mathcal{G}^{i}$, which ensures that the agent waits and sends a transaction order at the optimal time step. 

\vspace{2mm}

iv- \textit{Order book is filled and cleared}: A number $J$ of order books are filled with all the agents' trading limit orders for stock $j$ at time step $t$. All buy orders are sorted by descending bid prices, all sell orders are sorted by ascending ask prices, each with their own associated number of stocks to trade. Then the order book is cleared at this same time step $t$ by matching each order at mid-price between buyers and sellers, starting from the top of the order book to the lowest level where the bid price still exceeds the ask price. Importantly, we then define the market price $P^{j}(t+1)$ of stock $j$ at the next time step $t$ as that last and lowest level mid-price cleared by the order book. We also define the trading volume $V^{j}(t+1)$ as the number of stocks $j$ traded during that same time $t$. Finally, we define the spread $S^{j}(t+1)$ of stock $j$ at this time step $t$ as the absolute difference between the average of all bids and asks. Notice it is this spread $S^{j}(t)$ that is used as input to each agent's stock pricing process, and not the traditional bid-ask spread, which is defined as the difference between the highest bid and the lowest ask. 
\vspace{1mm}

\subsection{Agents parameters}

Let $\mathcal{U} (), \mathcal{U} \{ \}, \mathcal{N} (), \mathcal{N} \{ \}$ respectively denote the continuous and discrete uniform distributions, and the continuous and discrete normal distributions. Each agent $i$ is initialised at \textit{step $1$} with the following parameters : 
\begin{itemize}

\item[--] Risk-free assets of value $A_{\text{bonds}}^{i}(t=0) \sim \mathcal{N}(0,10^4)$. This can be seen as bonds or a bank account out of which the agent may long equity, and that will increase when the agent shorts its stocks. 

\item[--] A number of stocks $Q^{i,j}(t=0)$, drawn from a discrete positive half-normal distribution $\mathcal{N}^{+} \{ 0,100 \}$, amounting to a value of its stock holdings $A_{\text{equity}}^{i}(t=0)=\sum_{j=0}^JQ^{i,j}(t=0)P^{j}(t=0)$, which the agent may short on the market. 

\item[--] A drawdown limit $l^{i} \sim \mathcal{U} (50 \%, 60\%)$. If the peak-to-bottom percentage decrease since the beginning of the year in the agent's portfolio net asset value exceeds this drawdown limit $l^{i}$ at any time step $t$, then the agent is listed as bankrupt and unable to interact with the market anymore. 

\item[--] A reflexivity amplitude parameter $\rho^{i} \sim \mathcal{U} (0, 100\%)$, which gauges how fundamental or chartist the agent is in its price valuation, via a weighted average of the agent's technical forecast of the market price $\hat{P}^{i,j}(t)$ and its fundamental pricing $\mathcal{B}^{i,j}(t)$. The value of $\rho^{i}$ modulates the amplitude of the action $a_2^{\mathcal{F}}$ of the first reinforcement learning algorithm $\mathcal{F}$ (see below). 

\item[--] An investment horizon $\tau^{i} \sim \mathcal{U} \{T_w, 6T_m \}$, corresponding to the number of time steps after which the agent liquidates its position. Notice the bounds of this interval correspond to one week and six months in trading days. 

\item[--] A trading window $w^{i} \sim \mathcal{U} \{ T_w, \tau^{i} \}$, which will enter as parameter to the function $\mathcal{G}^{i}$ computing the optimal trading time for longing a certain quantity of stocks $j$. 

\item[--] A memory interval $h^{i} \sim \mathcal{U} \{ T_w, T-\tau^{i}-2T_w \}$, and corresponding to the fixed size of the rolling interval memorised by the agent for its learning process. 

\item[--] A transaction gesture $g^{i} \sim \mathcal{U} (0.2, 0.8)$, and related to how far above or below the value of its own stock pricing the agent is willing to trade and deal a transaction.  

\item[--] A reinforcement learning rate, which in our case is modelled by a parameter $\beta \sim \mathcal{U} (0.05, 0.20)$ for both reinforcement algorithms $\mathcal{F}^{i}$ and $\mathcal{T}^{i}$. 


\end{itemize}

\vspace{1mm}


\subsection{Agent reinforcement learning}

We now describe further \textit{step $3$} and its two reinforcement learning algorithms $\mathcal{F}^{i}$ (which learns efficient price forecasting) and $\mathcal{T}^{i}$ (which learns efficient trading based on this forecast). Each algorithm is individually ran by each agent $i$ following a direct policy search, for each stock $j$, and at each time step $t$. By direct policy search, we mean each agent selects and updates the probability associated with each action directly from the policy, and not via any action-value function (cf. GPI theorem~\cite{SuttonBarto}). Each algorithm has $729$ and $972$ potential action-state pairs, respectively. We define the sets of states $\mathcal{S}$, actions $\mathcal{A}$, and returns $\mathcal{R}$ of these two algorithms according to the following. 

\subsubsection{First algorithm $\mathcal{F}^{i}$}

Via this first algorithm, the agent continuously monitors the longer-term volatility of the stock prices ($s_0^{\mathcal{F}}$), their shorter-term volatility ($s_1^{\mathcal{F}}$), and the gap between its own present fundamental valuation and the present market price ($s_2^{\mathcal{F}}$). Out of this state, it learns to optimize its price prediction at its investment horizon $\tau^{i}$ by exploring and exploiting three possible actions via a direct policy search: choosing a simple forecasting econometric tool based on mean-reverting, averaging, or trend-following market prices ($a_0^{\mathcal{F}}$), choosing the size of the past time interval for this forecast ($a_1^{\mathcal{F}}$), and choosing the weight of its own fundamental stock pricing in an overall future price estimation, that is both fundamentalist and chartist ($a_2^{\mathcal{F}}$). 

\vspace{3mm}

\textit{States $\mathcal{S}^{\mathcal{F}}$}: The first reinforcement learning algorithm $\mathcal{F}^{i}$ is defined with a state $s^{\mathcal{F}}$ in a state set $\mathcal{S}^{\mathcal{F}}=\{ s_0^{\mathcal{F}}, s_1^{\mathcal{F}}, s_2^{\mathcal{F}} \}$, of dimension $27$, where each $s_0^{\mathcal{F}}, s_1^{\mathcal{F}}, s_2^{\mathcal{F}}$ individually may take the values $0$, $1$, or $2$. First, each agent computes the variances $\sigma_{L}$ and $\sigma_{S}$ of the prices $P^{j}(t)$, over the intervals $[t-3 \tau^{i}, t ]$ and $[t-\tau^{i}, t ]$, respectively. Then: 
\begin{itemize}
\item[--] The value $\sigma_{L}$ is computed and recorded at each time step in a time series that is continually sorted in ascending order and truncated to keep a size corresponding to agent memory interval $h^{i}$. The percentile of its present value at time step $t$ sets $s_0^{\mathcal{F}}=0$ if it is below $25\%$, $s_0^{\mathcal{F}}=2$ if it is above $75\%$, and $s_0^{\mathcal{F}}=1$ otherwise. The state parameter $s_0^{\mathcal{F}}$ thus gives the agent an idea of the longer-term volatility of the prices $P^{j}(t)$ of stock $j$, not in the sense of absolute static thresholds of this longer-term volatility, but of dynamic values depending on the agent's past history. 
\item[--] The value $\sigma_{S}$ is computed and recorded at each time step in a time series that is likewise continually sorted in ascending order and truncated to keep a size corresponding to agent memory interval $h^{i}$. The percentile of its present value at time step $t$ sets $s_1^{\mathcal{F}}=0$ if it is below $25\%$, $s_1^{\mathcal{F}}=2$ if it is above $75\%$, and $s_1^{\mathcal{F}}=1$ otherwise. The state parameter $s_1^{\mathcal{F}}$ thus gives the agent an idea of the shorter-term volatility of the prices $P^{j}(t)$ of stock $j$, not in the sense of absolute static thresholds of this shorter-term volatility, but of dynamic values depending on the agent's past history. Together, with the longer-term volatility, this shorter-term gives the agent a finer perception of the market price microstructure for its forecasting. The state parameter $s_2^{\mathcal{F}}$ thus gives the agent a sense of distance between present market price and its own fundamental valuation. 
\item[--] The average of $\lvert P^{j}(t)-\mathcal{B}^{i,j}(t) \rvert / P^{j}(t)$ is computed over the interval $[t-3 \tau^{i}, t ]$, and sets $s_2^{\mathcal{F}}=0$ if it is below $10\%$, $s_2^{\mathcal{F}}=2$ if it is above $30\%$, and $s_2^{\mathcal{F}}=1$ otherwise. 
\end{itemize}

\vspace{3mm}

\textit{Actions $\mathcal{A}^{\mathcal{F}}$}: Then, the reinforcement learning algorithm $\mathcal{F}^{i}$ is defined with an action $a^{\mathcal{F}}$ in an action set $\mathcal{A}^{\mathcal{F}}=\{ a_0^{\mathcal{F}}, a_1^{\mathcal{F}}, a_2^{\mathcal{F}} \}$, of dimension $27$, where each $a_0^{\mathcal{F}}, a_1^{\mathcal{F}}, a_2^{\mathcal{F}}$ individually may take the values $0$, $1$, or $2$. These actions are chosen according to a direct policy search (see below). First, each agent computes two different averages $\langle P^{j}_{[t-2T, t -T]}(t) \rangle$ and $\langle P^{j}_{[t-T, t]}(t) \rangle$ of past stock prices, with $T=(1+a_1^{\mathcal{F}}) \tau^{i}/2$, and then the econometric tool computes: 
\begin{eqnarray}
\hat{P}^{i,j}(t)&=&P^{j}(t)+\langle P^{j}_{[t-2T, t -T]}(t) \rangle - \langle P^{j}_{[t-T, t]}(t) \rangle  \nonumber \\
\hat{P}^{i,j}(t)&=& \frac{1}{2} \langle P^{j}_{[t-2T, t -T]}(t) \rangle + \frac{1}{2} \langle P^{j}_{[t-T, t]}(t) \rangle  \nonumber \\
\hat{P}^{i,j}(t)&=&P^{j}(t)-\langle P^{j}_{[t-2T, t -T]}(t) \rangle + \langle P^{j}_{[t-T, t]}(t) \rangle \nonumber
\end{eqnarray}

\noindent if $a_0^{\mathcal{F}}=0,1, 2$, respectively, and hence corresponding to mean-reverting, moving-average, and trend-following projections. Hence, both $a_0^{\mathcal{F}}$ and $a_1^{\mathcal{F}}$ pertain to technical analysis: action $a_0^{\mathcal{F}}$ determines the nature of the econometric forecasting tool that will be used (mean-reverting, moving-average, and trend-following), and action $a_1^{\mathcal{F}}$ determines the size of the past intervals over which these forecasts are computed. Then, the third action $a_2^{\mathcal{F}}$ enters as a parameter of the weighted average of this latter chosen technical forecast $\hat{P}^{i,j}(t)$ and the agent's fundamental valuation of the stock $\mathcal{B}^{i,j}(t)$, to produce the agent's forecast: 
\begin{eqnarray}
H^{i,j}(t)=\alpha \hat{P}^{i,j}(t) + (1-\alpha) \mathcal{B}^{i,j}(t)
\end{eqnarray}

\noindent for $\alpha \in \mathbb{R}$, which is specified such that if the agent's reflexivity $\rho^{i} \leq 50 \%$, we have $\alpha=0, \rho^{i}, 2\rho^{i}$ for  $a_2^{\mathcal{F}}=0,1, 2$, respectively, and if the agent's reflexivity $\rho^{i} > 50 \%$, we have $\alpha=2 \rho^{i}-1, \rho^{i}, 1$ for  $a_2^{\mathcal{F}}=0,1, 2$, respectively. Hence $a_2^{\mathcal{F}}=2$ allows the agent to learn and gauge the weight it gives to its own chartist vs. fundamentalist pricing. 

\vspace{3mm}

\textit{Returns $\mathcal{R}^{\mathcal{F}}$}: Then, the reinforcement learning algorithm $\mathcal{F}^{i}$ computes the percentage difference between the agent's former stock price prediction $H^{i,j}(t- \tau^{i})$ performed $\tau^{i}$ time steps ago, and its present realization $P^{j}(t)$:
\begin{eqnarray}
\frac{\lvert H^{i,j}(t- \tau^{i}) - P^{j}(t) \rvert }{ P^{j}(t)}
\end{eqnarray}

\noindent recording it at each time step in a time series that is continually sorted in ascending order and truncated to keep a size corresponding to agent memory interval $h^{i}$. The associated percentile corresponding to this value at time step $t$ sets a discrete value of returns $r^{\mathcal{F}}$ in the set $\mathcal{R}^{\mathcal{F}}=\{ 4,2,1,-1,-2,-4 \}$ if it respectively belongs to the intervals $[ 0\%, 5\%($, $[ 5\%, 25\%($, $[ 25\%, 50\%($, $[ 50\%, 75\%($, $[ 75\%, 95\%($, $[ 95\%, 100\%]$. 

\vspace{3mm}

\textit{Policy $\mathcal{\pi}^{\mathcal{F}}$}: Finally, the reinforcement learning algorithm updates its policy $\pi_t^{\mathcal{F}}(s^{\mathcal{F}}_{t- \tau^{i}},a^{\mathcal{F}}_{t- \tau^{i}})$, at each time step $t$. This is done according to the agent's own learning rate $\beta$, with the equations below being iterated a number $\lvert r^{\mathcal{F}} \rvert$ of times, in order to favour any action that is deemed optimal $a^{\mathcal{F}\star}$ in state $s^{\mathcal{F}}$, by increasing the policy probability associated with this action, compared to the other actions, $\forall a^{\mathcal{F}} \neq a^{\mathcal{F} \star}$ : 
\begin{eqnarray}
\pi^{\mathcal{F}}_{t+1} (s^{\mathcal{F}}, a^{\mathcal{F} \star}) &=& \pi^{\mathcal{F}}_t (s^{\mathcal{F}},a^{\mathcal{F} \star}) + \beta [ 1 - \pi^{\mathcal{F}}_t (s^{\mathcal{F}},a^{\mathcal{F} \star}) ] \nonumber \\
\pi^{\mathcal{F}}_{t+1} (s^{\mathcal{F}}, a^{\mathcal{F}}) &=& \pi^{\mathcal{F}}_t (s^{\mathcal{F}},a^{\mathcal{F}}) + \beta [ 0 - \pi^{\mathcal{F}}_t (s^{\mathcal{F}},a^{\mathcal{F}}) ] \nonumber
\end{eqnarray}

Added to this, the algorithm uses an off-policy method every $\tau^{i}/T_m + 2$ time steps, which computes the optimal action that $\mathcal{F}^{i}$ should have performed $\tau^{i}$ time steps ago now that the price is realised and the forecast accuracy known, and which accordingly updates the policy $\pi^{\mathcal{F}}$ with the agent's own learning rate $\beta$, iterated $\lvert r^{\mathcal{F}} \rvert=4$ times (for the associated action is deemed optimal).

\subsubsection{Second algorithm $\mathcal{T}^{i}$}

Via this second algorithm, the agent continuously monitors whether the stock prices are increasing or decreasing according to former algorithm ($s_0^{\mathcal{T}}$), their volatility ($s_1^{\mathcal{T}}$), its risk-free assets ($s_2^{\mathcal{T}}$), its quantity of stock holdings ($s_3^{\mathcal{T}}$), and the traded volumes ($s_4^{\mathcal{T}}$). Out of this state, it learns to optimize its investments by exploring and exploiting two possible actions via a direct policy search: sending a transaction order to the order book as holding, buying, or selling a position in a given amount ($a_0^{\mathcal{T}}$), and at what price wrt. the law of supply and demand ($a_1^{\mathcal{T}}$). 

\vspace{3mm}

\textit{States $\mathcal{S}^{\mathcal{T}}$}: The second reinforcement learning algorithm $\mathcal{T}^{i}$ is defined with a state $s^{\mathcal{T}}$ in a state set $\mathcal{S}^{\mathcal{T}}=\{ s_0^{\mathcal{T}}, s_1^{\mathcal{T}}, s_2^{\mathcal{T}}, s_3^{\mathcal{T}}, s_4^{\mathcal{T}} \}$, of dimension $108$, with $s_0^{\mathcal{T}}=s_1^{\mathcal{T}}=s_4^{\mathcal{T}}=\{ 0, 1, 2 \}$ and $s_2^{\mathcal{T}}=s_3^{\mathcal{T}}=\{ 0, 1 \}$. 
\begin{itemize} 

\item[--] Each agent computes the value $\mu=(H^{i,j}(t)-P^{j}(t))/P^{j}(t)$ and records it at each time step in two distinct time series $\boldsymbol{\mu}_{-}$ and $\boldsymbol{\mu}_{+}$, depending on whether it is negative or positive, respectively. These two time series are continually sorted in ascending order and truncated to keep a size corresponding to agent memory interval $h^{i}$. The percentile of its present value $\overline{\mu}_{-}$ at time step $t$ in $\boldsymbol{\mu}_{-}$ sets $s_0^{\mathcal{T}}=0$ if it is below $95\%$, and $s_0^{\mathcal{T}}=1$ otherwise. The percentile of its present value $\overline{\mu}_{+}$ at time step $t$ in $\boldsymbol{\mu}_{+}$ sets $s_0^{\mathcal{T}}=1$ if it is below $5\%$, and $s_0^{\mathcal{T}}=2$ otherwise. Therefore, $s_0^{\mathcal{T}}=0, 1, 2$ if the econometric forecast $\mu$ derived from the previous algorithm $\mathcal{F}^{i}$ is respectively indicating a decrease, approximate stability, or increase in the price of stock $j$ in $\tau^{i}$ future time steps. 

\item[--] Each agent records the previously computed variance $\sigma_{L}$ of the prices $P^{j}(t)$ over the interval $[t-3 \tau^{i}, t ]$ at each time step, in a time series that is continually sorted in ascending order and truncated to keep a size corresponding to agent memory interval $h^{i}$. The percentile of its present value at time step $t$ sets $s_1^{\mathcal{T}}=0$ if it is below $33\%$, $s_1^{\mathcal{T}}=2$ if it is above $67\%$, and $s_1^{\mathcal{T}}=1$ otherwise. $s_1^{\mathcal{T}}$ thus gives a measure of the longer-term volatility in stock prices to the agent. 

\item[--] Each agent sets $s_2^{\mathcal{T}}=0$ if the value of its risk-free assets $A_{\text{bonds}}^{i}(t)$ at time step $t$ is below $60 \%$ of its start value $A_{\text{bonds}}^{i}(t=0)$, and $s_2^{\mathcal{T}}=1$ otherwise. Hence each agent likewise continually monitors the size of its risk-free assets in order to adopt the appropriate long or short strategy. 

\item[--] Each agent sets $s_3^{\mathcal{T}}=0$ if the value of its stock holdings $A_{\text{equity}}^{i}(t)$ at time step $t$ is below $60 \%$ of its start value $A_{\text{equity}}^{i}(t=0)$, and $s_3^{\mathcal{T}}=1$ otherwise. Hence each agent continually monitors the size of its stock holdings in order to adopt the appropriate long or short strategy. 

\item[--] Each agent records the trading volume $V^{j}(t)$ at each time step in a time series that is continually sorted in ascending order and truncated to keep a size corresponding to agent memory interval $h^{i}$. The percentile of its present value at time step $t$ sets $s_4^{\mathcal{T}}=0$ if $V^{j}(t)=0$, $s_4^{\mathcal{T}}=1$ if it is below $33\%$, $s_4^{\mathcal{T}}=2$ otherwise. This gives the agent a sense of market activity that is useful for determining the appropriate bid or ask price to send to the order book. 
\end{itemize}

\vspace{3mm}

\textit{Actions $\mathcal{A}^{\mathcal{T}}$}: Then, the reinforcement learning algorithm $\mathcal{T}^{i}$ is defined with an action $a^{\mathcal{T}}$ in an action set $\mathcal{A}^{\mathcal{T}}=\{ a_0^{\mathcal{T}}, a_1^{\mathcal{T}} \}$, of dimension $9$. Here $a_0^{\mathcal{T}}$ and $a_1^{\mathcal{T}}$ both can take the discrete values $\{ 0, 1, 2 \}$, chosen according to a direct policy search (see below). The first action $a_0^{\mathcal{T}}$ corresponds both to the quantity of stocks and the nature of the transaction order (sell, hold, or buy) that the agent chooses to send to the order book. The second action $a_1^{\mathcal{T}}$ corresponds to the flexibility of the agent with regards to the price at which it is willing to trade. As said, these two actions depend on the evaluation of the price of stock $j$ that was performed by the agent through the first algorithm $\mathcal{F}^{i}$. First, the agent bid price $P^{i,j}_{\text{bid}}(t)$ is set at:
\begin{eqnarray}
P^{i,j}_{\text{bid}}(t) &=& \min [H^{i,j}(t), P^{j}(t)] +  g^{i}S^{j}(t-1) \nonumber \\
P^{i,j}_{\text{bid}}(t) &=& \min [H^{i,j}(t), P^{j}(t)] \nonumber \\
P^{i,j}_{\text{bid}}(t) &=& \min [H^{i,j}(t), P^{j}(t)] -  g^{i}S^{j}(t-1)  \nonumber 
\end{eqnarray}

\noindent if $a_1^{\mathcal{T}}={0,1,2}$ respectively. Here we recall that $g^{i}$ is the agent's transaction gesture and $S^{j}(t-1)$ our market spread of stock $j$ at former time step. The term $\pm g^{i}S^{j}(t-1)$ hence specifies the agent's softer or harder stance on the transaction deal, depending on general market conditions of supply and demand, like $S^{j}(t-1)$ and the traded volumes specified by $s_4^{\mathcal{T}}$. The agent ask price $P^{i,j}_{\text{ask}}(t)$ is set at: 
\begin{eqnarray}
P^{i,j}_{\text{ask}}(t) &=& \max [H^{i,j}(t), P^{j}(t)] -  g^{i}S^{j}(t-1) \nonumber \\
P^{i,j}_{\text{ask}}(t) &=& \max [H^{i,j}(t), P^{j}(t)] \nonumber \\
P^{i,j}_{\text{ask}}(t) &=& \max [H^{i,j}(t), P^{j}(t)] +  g^{i}S^{j}(t-1) \nonumber 
\end{eqnarray}

\noindent if $a_1^{\mathcal{T}}={0,1,2}$ respectively. Then for $Q^{i,j}(t)$ the quantity of stocks $j$ held by agent $i$ at time $t$, action $a_0^{\mathcal{T}}=0$ corresponds to a sell order of a quantity $Q^{i,j}(t)$ of stocks $j$ for an ask price of $P^{i,j}_{\text{ask}}(t)$, action $a_0^{\mathcal{T}}=1$ corresponds to no order being sent to the order book (the agent simply holds its position), and $a_0^{\mathcal{T}}=2$ corresponds to a buy order of a floored quantity $A_{\text{bonds}}^{i}(t)/[P^{i,j}_{\text{ask}}(t)J]$ of stocks $j$ for a bid price of $P^{i,j}_{\text{bid}}(t)$. Notice the parameter $J$ is here part of the denominator of this stock quantity to buy, so as to ensure a proper multivariate portfolio management. 

\vspace{3mm}

\textit{Filter $\mathcal{G}^{i}$}: As mentioned earlier, the quantity and price of stock $j$ that agent $i$ sends to the order book at this time step $t$ is conditional on the output of function $\mathcal{G}^{i}$, that ensures the agent waits and sends a transaction order at the optimal time step and not before. In order to do this, $\mathcal{G}^{i}$ records in a time series at each time step the value of the action-value function $\arg \max_{a} \mathcal{Q}_{t}(s,a)$ maximized by action $a$. It then sorts this time series in ascending order, and compares its associated percentile $p_{\mathcal{Q}}(t)$ with the ratio of the elapsed time since previous transaction $k^{i,j}(t)$ over the agent's individual trading window $w^{i}$. The filter function $\mathcal{G}^{i}$ lets the trading order chosen by $\mathcal{T}^{i}$ be sent to the order book only if $p_{\mathcal{Q}}(t)<k^{i,j}(t)/w^{i}$. Notice this function $\mathcal{G}^{i}$ thus filters entry but not exit strategies: the latter are always enacted at the agent's investment horizon $\tau^{i}$. 

\vspace{3mm}

\textit{Returns $\mathcal{R}^{\mathcal{T}}$}: Considering the present realization of stock price $P^{j}(t)$, the algorithm $\mathcal{T}^{i}$ then computes the cashflow difference between the present agent's portfolio net asset value, and its present value if the former actions taken $\tau^{i}$ time steps ago had not been taken:
\begin{eqnarray}
Q^{i,j}_{\text{OB}}(t-\tau^{i}) [P^{j}(t) - P^{i,j}_{\text{OB}}(t-\tau^{i})]
\end{eqnarray}

\noindent where $Q^{i,j}_{\text{OB}}(t-\tau^{i})$ and $P^{i,j}_{\text{OB}}(t-\tau^{i})$ are respectively the quantity and transaction price of stock $j$ that was cleared by the order book process at time $t-\tau^{i}$ for agent $i$ and its transaction partner. Notice these may not be those actually sent by agent $i$ at that time, because the quantity of stocks to long or short may not have been entirely cleared at this time (recall the agents send limit orders only), and because the transaction price is set by the order book at mid-price with the transaction partner's order price. These values are then recorded at each time step in a time series that is continually sorted in ascending order and truncated to keep a size corresponding to agent memory interval $h^{i}$. The associated percentile corresponding to this value at time step $t$ sets a discrete value of returns $r^{\mathcal{T}}$ in the set $\mathcal{R}^{\mathcal{T}}=\{ 4,2,1,-1,-2,-4 \}$ if it respectively belongs to the intervals $[ 0\%, 5\%($, $[ 5\%, 25\%($, $[ 25\%, 50\%($, $[ 50\%, 75\%($, $[ 75\%, 95\%($, $[ 95\%, 100\%]$. 

\vspace{3mm}

\textit{Policy $\mathcal{\pi}^{\mathcal{T}}$}: Finally, the reinforcement learning algorithm updates its policy $\pi_t^{\mathcal{T}}(s^{\mathcal{T}}_{t- \tau^{i}},a^{\mathcal{T}}_{t- \tau^{i}})$, every $\tau^{i}$ time steps following each transaction dealt by the agent. This is done according to the agent's own learning rate $\beta$, with the equations below being iterated a number $\lvert r^{\mathcal{T}} \rvert$ of times, in order to favour any action that is deemed optimal $a^{\mathcal{T}\star}$ in state $s^{\mathcal{T}}$, by increasing the policy probability associated with this action, compared to the other actions, $\forall a^{\mathcal{T}} \neq a^{\mathcal{T} \star}$ : 
\begin{eqnarray}
\pi^{\mathcal{T}}_{t+1} (s^{\mathcal{T}}, a^{\mathcal{T} \star}) &=& \pi^{\mathcal{T}}_t (s^{\mathcal{T}},a^{\mathcal{T} \star}) + \beta [ 1 - \pi^{\mathcal{T}}_t (s^{\mathcal{T}},a^{\mathcal{T} \star}) ] \nonumber \\
\pi^{\mathcal{T}}_{t+1} (s^{\mathcal{T}}, a^{\mathcal{T}}) &=& \pi^{\mathcal{T}}_t (s^{\mathcal{T}},a^{\mathcal{T}}) + \beta [ 0 - \pi^{\mathcal{T}}_t (s^{\mathcal{T}},a^{\mathcal{T}}) ] \nonumber
\end{eqnarray}

Added to this, the algorithm uses an off-policy method every $\tau^{i}/T_m + 2$ time steps, which computes the optimal action that $\mathcal{T}^{i}$ should have performed $\tau^{i}$ time steps ago now that the price is realised and the forecast accuracy known, and which accordingly updates the policy $\pi^{\mathcal{T}}$ with the agent's own learning rate $\beta$, iterated $\lvert r^{\mathcal{T}} \rvert=4$ times (for the associated action is deemed optimal). 

\vspace{1mm}

As one can see, the action-state spaces of these two algorithms $\mathcal{F}$ and $\mathcal{T}$ are highly discretised and handcrafted. A main reason for this is that a central challenge of financial MAS research has historically been the large computational power required by such models. Furthermore, the general intuition behind our definition of such state and action spaces has been the \textit{Fundamental Theorem of Asset Pricing}~\cite{Delbaen2011}, and these are not that dissimilar from other recent models such as~\cite{Spooner2018} (cf. section 4.3).

\vspace{1mm}

\section{Calibration}

\textit{Model hypotheses}: The main hypotheses for this model are that: i- the simulated agents faithfully emulate real-world investors, and ii- the simulated transaction limit orders processed by the order book correspond to the dynamics and properties of real stock market orders. Concerning the former, the strength of our approach is the simplification that any agent, regardless of its behaviour and strategy, is structurally bound to interact with the market in three possible ways only, namely by longing, shorting or holding stocks. Concerning the latter, we recall that the dynamics of order books are extensively documented in literature~\citep{Huang2015}, and hence that their design can be rigorously conducted. 

\vspace{3mm}

\textit{Model limitations}: With these two major hypotheses, we shall also mention the following model limitations and consistency issues, which are proper to all financial MAS: i- reliance on virtual fundamental generation $\mathcal{T}^{j}(t)$, ii- absence of portfolio diversification (equity, cross-asset), iii- absence of various trading strategies (e.g. short-selling, leveraging, derivatives, metaorders, market orders, etc.), iv- absence of intraday and seasonal market effects, v- absence of legal and regulatory constraints. Although some of these limitations may seem challenging, their effects and importance virtually impact any other econometric or modelling approach in quantitative finance. Added to this, modelling market activity via a market microstructure emerging from a centralised order book that processes transaction orders sent by many trading agents, has a direct empirical correspondance with real stock markets, and is thus fully epistemologically pertinent. 

\vspace{3mm}

\textit{Training and testing sets}: We calibrated the MAS stock market simulator to real stock market data~\footnote{Computations were performed on a Mac Pro with 3,5 GHz 6-Core Intel Xeon E5 processor, and 16 GB 1866 MHz DDR memory}. In order to do so, we used high quality, industry-grade, daily close prices and volumes of $4,313$ stocks from the London Stock Exchange (LSE), between January $15$th of 2007 and January $19$th of 2018. In order to work on the market microstructure, we have filtered the data as such: i- stock-splits effects have been suppressed, ii- only stocks that have been continuously traded over this time period have been considered. As a consequence of this data curation, our former stock universe has been lowered to $640$ stocks. The calibration of the MAS hyperparameters has been conducted on a random sampling of half of these stocks as training set, whose statistical features are virtually identical to the other half. We posit this statistical stability, proper to stock market data, arises from the lack of market arbitrage and relates to the aforementioned \textit{stylized facts}. 

\vspace{3mm}

\textit{Hyperparameter optimisation}: The calibrated hyperparameters are the number of agents $I$, the agent transaction gesture factor $\zeta^i \in \mathbb N$ (which scales the gesture parameter $g^i$ initialised for each agent at time $t=0$), the fundamental values generation parameter $\nu$ (which is the amplitude of the fundamental time series $\mathcal{T}^j (t)$), and the drawdown threshold (which is added to the drawdown limit $l^{i}$ initialised at time $t=0$ for each agent). All hyperparameter combinations were ran on the training set, and are listed on Tab. \ref{T1}.
\begin{table}[h!]
\centering
\begin{tabular}{ | m{4cm} | m{1cm}| m{1cm} | m{1cm} | } 
\hline
\textbf{Hyperparameter} & \textbf{Low} & \textbf{High} & \textbf{Step} \\ 
\hline
Agent number $I$ & $500$ & $5000$  & $500$ \\ 
\hline
Gesture scalar $\zeta^i$ & $1.0$ & $3.0$ & $0.5$ \\ 
\hline
Fundamental amplitude $\nu$ & $0.1$ & $1.5$ & $0.2$ \\ 
\hline
Drawdown treshold $\mathcal L$ & $-50$ & $30$ & $10$ \\ 
\hline
\end{tabular}
\caption{Model hyperparameters, with their intervals for training : lower bound (Low), higher bound (High), and step of incrementation (Step).}
\label{T1}
\end{table}

\vspace{3mm}

\begin{figure}[!htbp]
\begin{centering}
\includegraphics[scale=0.4]{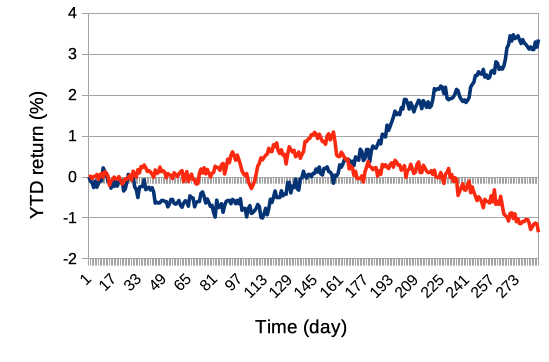}
\includegraphics[scale=0.4]{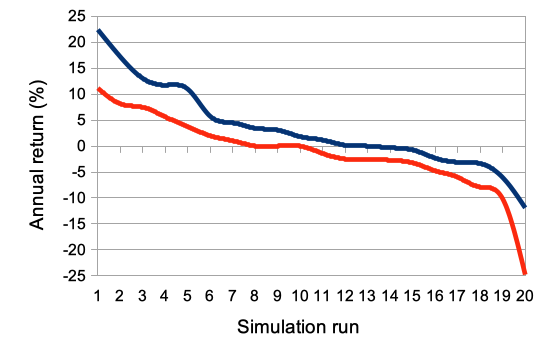}
\caption{\label{A13} After $90 \%$ of total simulation time, we want to compare the best $10\%$ agents of our MAS stock market simulator (blue curves) with the best $10\%$ agents of a market simulated with noise agents trading randomly (red curves). For this, we check on their performance over the remainder $10 \%$ of our total simulation, with averaged equity curves as their year-to-date returns over $S=20$ simulations (left), and averaged sorted annual returns of each $S$ simulations (right). These simulations are generated with parameters $I=500$, $J=1$, $T=2875$.}
\end{centering}
\end{figure}

\textit{Model comparison}: There is a huge literature wrt. to~\cite{Gode1993} on the substitution of markets to individual rationality, namely whether markets eliminate irrational individuals, or whether individuals learn market rules. We here provide on Fig. \ref{A13} agent learning curves. These can be used for model comparison, notably with recent order book models coupled with reinforcement learning~\cite{Spooner2018}, and the former generation of MAS with zero-intelligence agents~\cite{Gode1993} as baselines. 
\vspace{3mm}



\section{Performance}

\subsection{Statistical features}

\begin{figure}[!htbp]
\begin{centering}
\includegraphics[scale=0.53]{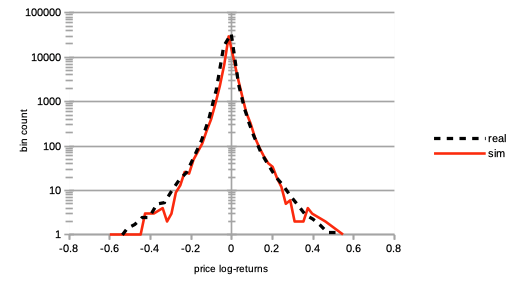}
\caption{\label{A1} Distribution of logarithmic returns of prices $\log [P(t)/P(t-1)]$ of real (dashed black curve) and simulated (continuous red curve) data. The simulations are generated with parameters $I=500$, $J=1$, $T=2875$, and $S=20$.}
\end{centering}
\end{figure}

\begin{figure}[!htbp]
\begin{centering}
\includegraphics[scale=0.53]{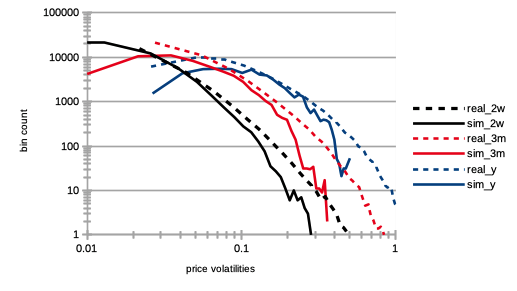}
\caption{\label{A2} Distribution of volatilities (defined as standard deviations of price normalised to price itself $\sigma/P(t)$) computed over lags of two weeks (black), three months (red), and one year (blue) intervals for both real (dashed curves) and simulated (continuous curves) data. The simulations are generated with parameters $I=500$, $J=1$, $T=2875$, and $S=20$.}
\end{centering}
\end{figure}

\begin{figure}[!htbp]
\begin{centering}
\includegraphics[scale=0.53]{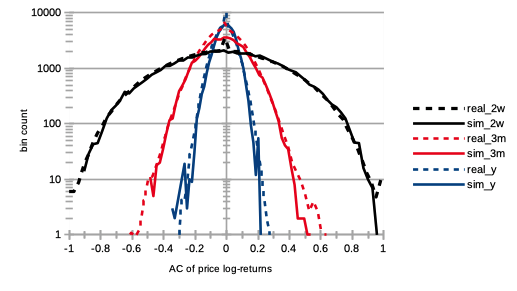}
\caption{\label{A3} Distribution of autocorrelations of the logarithmic returns of prices at each time step $t$ between intervals $[t-\Delta, t]$ and $[t-2\Delta, t-\Delta]$, over lags $\Delta$ of two weeks (black), three months (red), and one year (blue) intervals for both real (dashed curves) and simulated (continuous curves) data. The simulations are generated with parameters $I=500$, $J=1$, $T=2875$, and $S=20$.}
\end{centering}
\end{figure}

\begin{figure}[!htbp]
\begin{centering}
\includegraphics[scale=0.53]{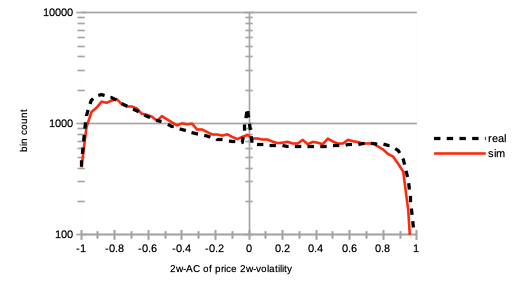}
\caption{\label{A4} Distribution of autocorrelations of two weeks-interval volatilities at each time step $t$ between intervals $[t-\Delta, t]$ and $[t-2\Delta, t-\Delta]$ for $\Delta=2T_{w}$, for both real (dashed black curve) and simulated (continuous red curve) data. The simulations are generated with parameters $I=500$, $J=1$, $T=2875$, and $S=20$.}
\end{centering}
\end{figure}

\begin{figure}[!htbp]
\begin{centering}
\includegraphics[scale=0.53]{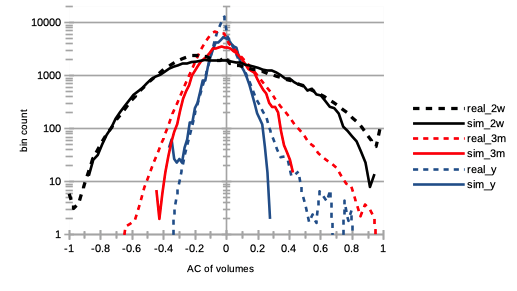}
\caption{\label{A5} Distribution of autocorrelations of the trading volumes at each time step $t$ between intervals $[t-\Delta, t]$ and $[t-2\Delta, t-\Delta]$, over lags $\Delta$ of two weeks (black), three months (red), and one year (blue) intervals for both real (dashed curves) and simulated (continuous curves) data. The simulations are generated with parameters $I=500$, $J=1$, $T=2875$, and $S=20$.}
\end{centering}
\end{figure}

\begin{figure}[!htbp]
\begin{centering}
\includegraphics[scale=0.53]{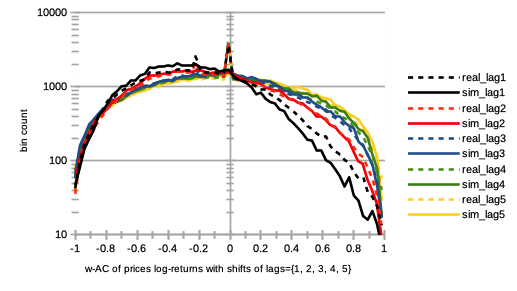}
\caption{\label{A6} Distribution of autocorrelations of the logarithmic returns of prices at each time step $t$ between intervals $[t-T_w, t]$ and $[t-T_w-\delta, t-\delta]$, for shifts $\delta$ of one day (black), two days (red), three days (blue), four days (green), and five days (yellow). This is for both real (dashed curves) and simulated (continuous curves) data. The simulations are generated with parameters $I=500$, $J=1$, $T=2875$, and $S=20$.}
\end{centering}
\end{figure}

\begin{figure}[!htbp]
\begin{centering}
\includegraphics[scale=0.53]{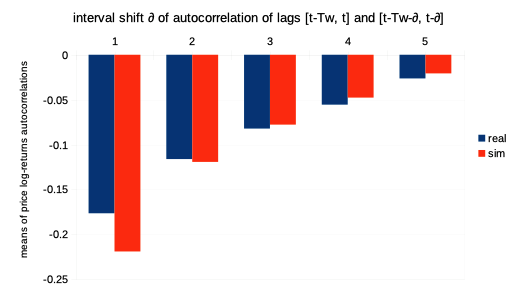}
\caption{\label{A7} Means of autocorrelations of the logarithmic returns of prices at each time step $t$ between intervals $[t-T_w, t]$ and $[t-T_w-\partial, t-\partial]$, for shifts $\partial=[1, 2, 3, 4, 5]$. This is for both real (blue) and simulated (red) data. The simulations are generated with parameters $I=500$, $J=1$, $T=2875$, and $S=20$.}
\end{centering}
\end{figure}

\begin{figure}[!htbp]
\begin{centering}
\includegraphics[scale=0.53]{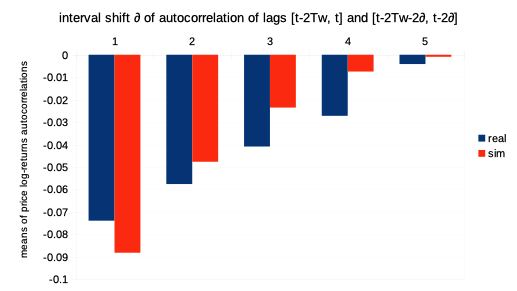}
\caption{\label{A8} Means of autocorrelations of the logarithmic returns of prices at each time step $t$ between intervals $[t-2T_w, t]$ and $[t-2T_w-\partial, t-\partial]$, for shifts $\partial=[2, 4, 6, 8, 10]$. This is for both real (blue) and simulated (red) data. The simulations are generated with parameters $I=500$, $J=1$, $T=2875$, and $S=20$.}
\end{centering}
\end{figure}

\begin{figure}[!htbp]
\begin{centering}
\includegraphics[scale=0.53]{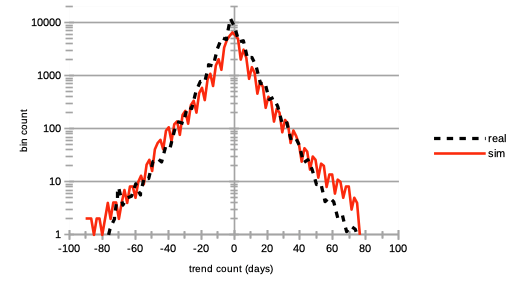}
\caption{\label{A9} Distribution of the number of consecutive days of increasing prices (positive values) and decreasing prices (negative values). This is for both real (dashed black curve) and simulated (continuous red curve) data. The simulations are generated with parameters $I=500$, $J=1$, $T=2875$, and $S=20$.}
\end{centering}
\end{figure}

We show on Fig. \ref{A1} to \ref{A12}, different key market microstructure indicators pertaining to the calibration of the MAS simulator. As one can see, the fits are satisfactory in topology compared to real stock market data, and constitute a real leap forward in terms of realism for MAS stock market simulators. This also shows the relevance of reinforcement learning as a framework for the description of agent learning and trading process in stock markets. 
\begin{itemize}
\item[--] On Fig. \ref{A1}, we show a rather great fit of logarithmic price returns to real stock market data. 

\item[--] On Fig. \ref{A2}, we plotted the price volatilities of both simulated and real data, over several intervals, namely $2T_w, 3T_m, T_y$. We found that the real volatilities at larger time scales are harder to emulate, and this is most probably because our real data sample covers a rather unique and unusual market regime over the years $2008-2010$, namely the financial crisis. 

\item[--] On Fig. \ref{A3}, we show the autocorrelations of the price logarithmic returns for both real and simulated data, over several intervals of length $2T_w, 3T_m, T_y$. Despite the great fits, one could wonder as to why real data shows such a disruptive shape at zero-autocorrelations: this could be due to intraday market activity, or to the absence of trading volumes over long periods of time for stocks of small capitalisation companies. 

\item[--] In a similar way, we see the simulated data emulating the real data in the asymmetric shapes of its distributions of autocorrelation in volatility (Fig. \ref{A4}) and volumes (Fig. \ref{A5}), and of blended autocorrelation in price logarithmic returns (Fig. \ref{A6}). 

\item[--] On Fig. \ref{A7} and \ref{A8}, we see the fits of simulated to real data, in the decrease of blended autocorrelations when parameter $\partial$ varies, for intervals of lengths $T_w$ and $2T_w$, respectively. This is very important wrt. the fact that our MAS produces a price microstructure that shows cancellation of arbitrage opportunities and market memory.

\item[--] On Fig. \ref{A9}, we show the distribution of the number of consecutive days of increasing prices (positive values) and decreasing prices (negative values) at each time step $t$, for both simulated and real data. The number of consecutive days of increasing or decreasing prices is a useful indicator of market regime, or whether the market is bearish or bullish. We find that apart from a few extreme bullish events, the MAS simulates general stock market price dynamics in a way corresponding to real data. 

\end{itemize}

\vspace{2mm}

In conclusion, we shall highlight the accurate emulation of the simulator with regards to the distribution of logarithmic price returns in Fig. \ref{A1}, their autocorrelations at different timescales in Fig. \ref{A3}, \ref{A6}, \ref{A7}, \ref{A8}. These latter autocorrelation metrics are extremely important in the calibration process, because they pertain to the absence of arbitrage and market memory, which are central features of financial markets. Or in other words, the simulated data like the real data should not display price patterns that are too easily identifiable and ready to be exploited by trading. We can also underline a rather remarkable emulation in market regimes, as shown in Fig. \ref{A9}. We also highlight some improvement left for:  

\begin{itemize}
\item[--] The extremity of the tail distribution of long-term price volatilities in Fig. \ref{A2}: these are indeed the hardest microstructure effects to capture, as they relate to jump diffusion processes proper to volatile events in the life of a company, industry sector, or full market (here we should mention this LSE data encompasses the financial crisis of $2008-2009$). 
\item[--] The peak in zero autocorrelations for real price returns and volatility in Fig. \ref{A3}-\ref{A4}: we posit this to be due to the fact that unlike real data, the simulator does not capture intraday market activity, or to the presence of scarcely traded small cap companies. 
\item[--] Fatter tails in distributions of autocorrelation of trading volumes in Fig. \ref{A5}: we posit this to be due to seasonal and calendar effects proper to real stock markets. 
\end{itemize}

\subsection{Classification features}

\begin{figure}[!htbp]
\begin{centering}
\includegraphics[scale=0.34]{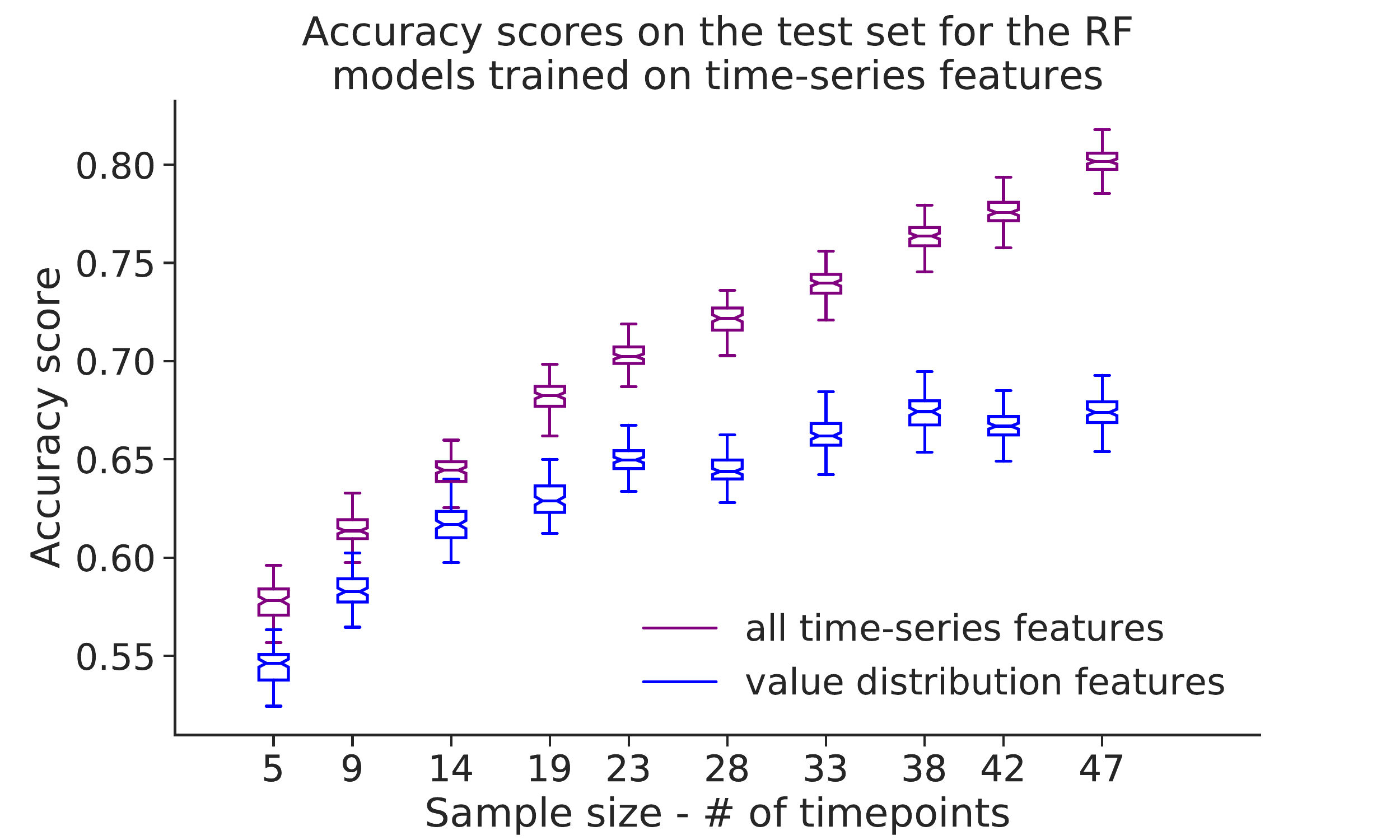}
\caption{\label{A10} Accuracy as a function of sample size. The value distribution features include only features that do not depend on the order of values in the time-series (e.g. mean, median, variance, skew, etc.), whereas all time-series features correspond to the total set of time-series features including those that depend on the temporal structure of the series (e.g. entropy, DFT coefficients, etc.). The simulations are generated with parameters $I=500$, $J=1$, $T=2875$, and $S=20$.}
\end{centering}
\end{figure}

\textit{Random forest classification}: In order to evaluate how well the model-generated log-return signals match the real market stock dynamics, we aimed to build a binary classification model which could take a fixed-size time series sample, and predict whether the sample comes from simulated or real stock market data. The whole dataset was split into non-overlapping training and validation subsets. Then, the time series samples of fixed size (the number of timestamps ranging from $5$ to $50$, but fixed for each experiment) were generated by applying a sliding window to full price logarithmic-return time series from both simulated and real data. Both training and validation datasets were balanced by construction in terms of class distribution. Before training a classifier on the collected data, a preprocessing step was taken to map the time series samples to a vector feature space by computing a set of $787$ predefined signal properties for each sample~\footnote{We used the implemented \textit{tsfresh} Python package \cite{tsfresh}}. A standard scaling procedure was then applied to the dataset with mean and variance parameters estimated from the training subset. This normalised feature-vector representation of the dataset was then used to train random forest classification models~\footnote{We used the implementation from \textit{scikit-learn} Python package \cite{sklearn}, with $200$ estimators and default values for other hyperparameters.}. Accuracy scores were measured for multiple runs of model training (with different random seed values) and for the different number of timestamps initially taken for each sample. The obtained accuracy score value distributions are shown in Fig. \ref{A10}. 

\begin{figure}[!htbp]
\begin{centering}
\includegraphics[scale=0.4]{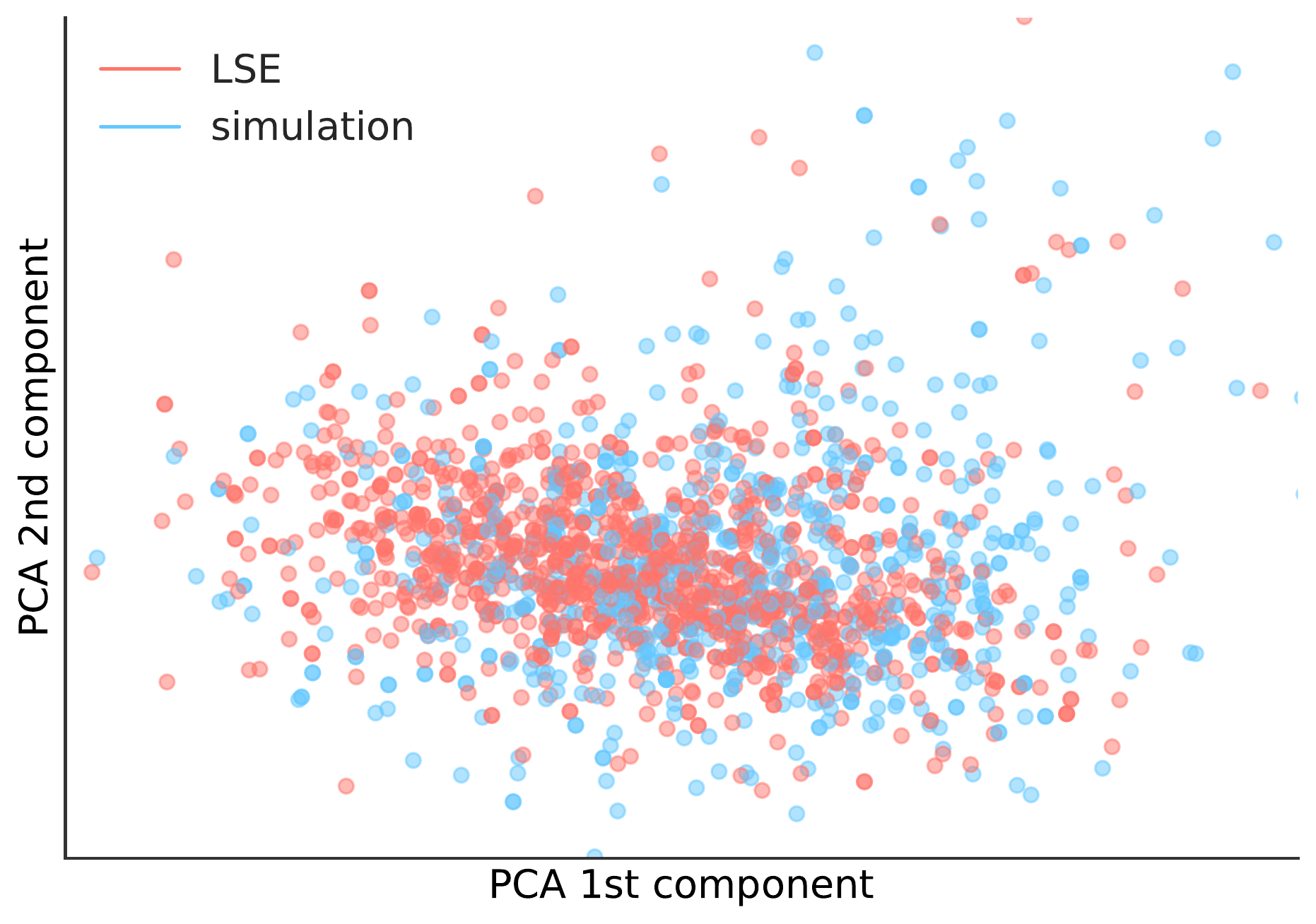}
\caption{\label{A11} Principal component analysis scatter plot on top $10$ statistical features of real (red) and simulation (blue). The simulations are generated with parameters $I=500$, $J=1$, $T=2875$, and $S=20$.}
\end{centering}
\end{figure}

\begin{figure}[!htbp]
\begin{centering}
\includegraphics[scale=0.4]{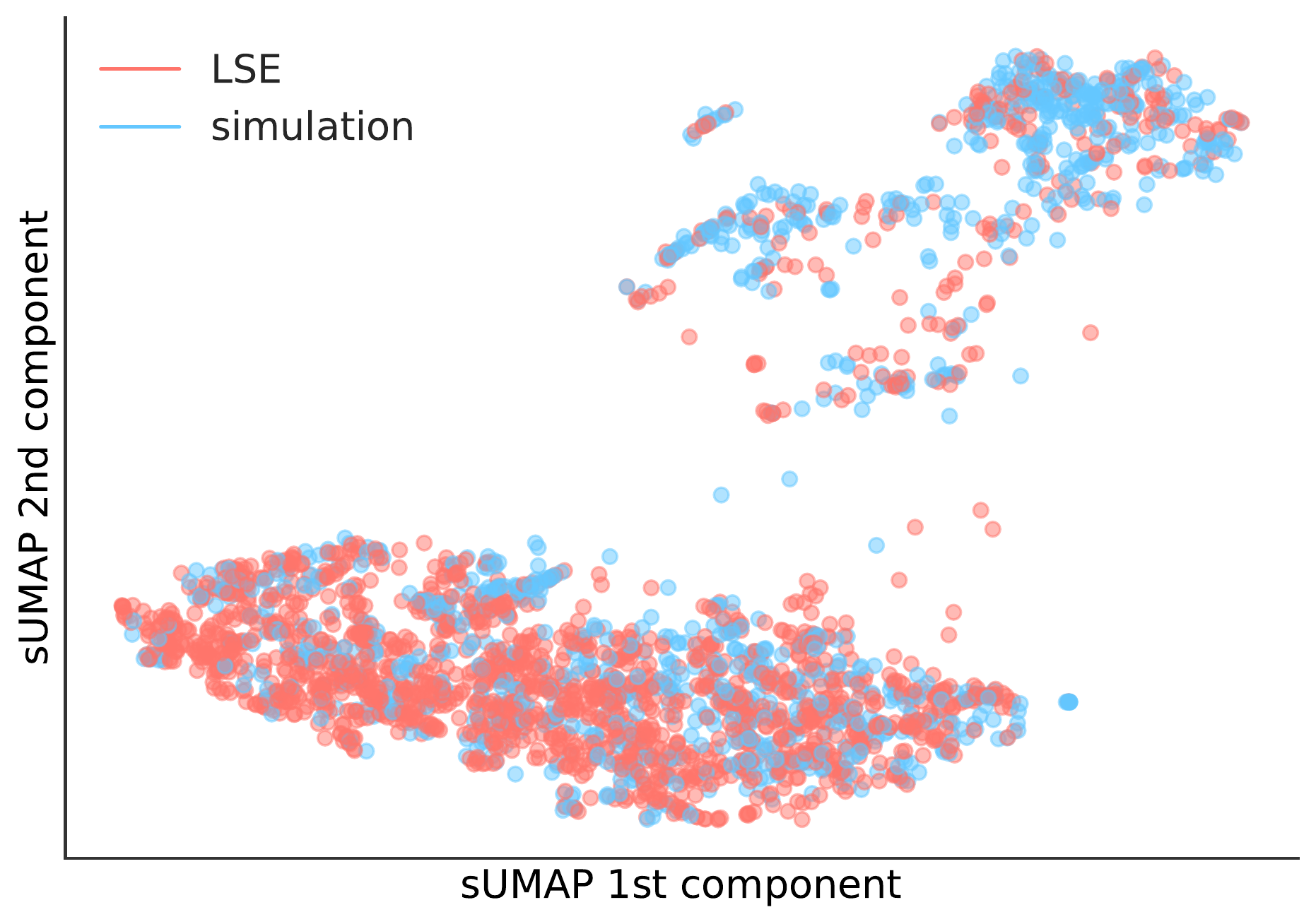}
\caption{\label{A12} Supervised UMAP trained on $2000$ samples for top $10$ statistical features of real (red) and simulation (blue), and applied to another $2000$ samples. The simulations are generated with parameters $I=500$, $J=1$, $T=2875$, and $S=20$.}
\end{centering}
\end{figure}

\textit{Dimensionality reduction}: Another useful metric that was extracted from the trained random forest classifiers is the measure of feature importance that was averaged over multiple training runs. This procedure allowed us to determine a subset of features which turn out as the most discriminative for the defined classification task. The features which were found in the top-$20$ importance score list could be grouped into several categories: i. value distribution properties (e.g. median value of the sample, $60$th, $70$th, $80$th percentiles of the value distribution within the sample), ii. fast Fourier transform spectrum statistics (centroid, skew and kurtosis of the fast Fourier transform coefficient distributions), iii. features related to the temporal structure of the series, e.g. autocorrelation and coefficients of a fitted autoregressive $AR(k)$ process. In order to visualise the distribution of classes across time series samples in the feature-vector space, we applied several dimensionality reduction algorithms to the dataset, results of which are shown in Fig. \ref{A11} and \ref{A12}. The first algorithm is generic principal component analysis, which was applied to the reduced set of top-$10$ features, ranked by feature importance scores, as seen on Fig. \ref{A11}. One can notice that the produced linear mapping to the 2-dimensional space does not guarantee high separability of point clouds corresponding to different classes (simulated vs. real data). This means that a linear combination of basic time-series features is not sufficient to effectively discriminate simulated data from the real signals. However, once we applied a supervised nonlinear mapping, namely Uniform Manifold APproximation (UMAP) \cite{umap}, class separability between point clouds increased significantly. Hence, a classifier learning nonlinear features from given time-series data (e.g. a deep convolutional neural network) can, in principle, achieve high classification quality metric values in this task. These considerations open further directions for model tuning/calibration, for instance by imposing a loss term which would depend on classification accuracy achieved by a deep neural network (e.g. as in Generative Adversarial Networks \cite{GANs}) in order to minimise it.

\vspace{3mm}


\section{Conclusion}

We thus modelled a stock market via an intelligent MAS, where the agents autonomously perform portfolio management via long-only equity strategies, based on autonomous reinforcement learning algorithms performing price forecasting and stock trading. In such a model, each agent also learns to gauge how fundamentalist or chartist it will be in its approach to price estimation. We have calibrated this MAS simulator to real stock market data from the London Stock Exchange between the years $2008$ and $2018$, an achieved state-of-the-art performance in price microstructure emulation. We posit this model could be a powerful tool for both financial industry and academic research, to further explore stock market microstructure and price formation by a bottom-up approach. Also, we posit that the reinforcement learning framework of the agents could be used to implement psychological traits of decision theory and behavioural economics, and hence to study the influence of agent learning and cognition on financial markets at the macroscale. Finally, we also see the following immediate and natural extensions of our model: i- using the multivariate feature of the agents trading coupled with portfolio risk management in order to study and simulate covariance structure across several stocks, and ii- extension of the agent reinforcement learning framework to the continuous domain (cf. policy gradients, DQN, etc.). 

\vspace{3mm}

\section{Acknowledgements}
We graciously acknowledge this work was supported by the RFFI grant nr. 16-51-150007 and CNRS PRC nr. 151199. I. L.'s work was supported by the Russian Science Foundation, grant nr. 18-11-00294.

\medskip

\bibliography{Article}

\begin{thebibliography}{101}
\expandafter\ifx\csname natexlab\endcsname\relax\def\natexlab#1{#1}\fi
\expandafter\ifx\csname bibnamefont\endcsname\relax
  \def\bibnamefont#1{#1}\fi
\expandafter\ifx\csname bibfnamefont\endcsname\relax
  \def\bibfnamefont#1{#1}\fi
\expandafter\ifx\csname citenamefont\endcsname\relax
  \def\citenamefont#1{#1}\fi
\expandafter\ifx\csname url\endcsname\relax
  \def\url#1{\texttt{#1}}\fi
\expandafter\ifx\csname urlprefix\endcsname\relax\def\urlprefix{URL }\fi
\providecommand{\bibinfo}[2]{#2}
\providecommand{\eprint}[2][]{\url{#2}}

\bibitem[{\citenamefont{Greene}(2017)}]{Greene2017}
\bibinfo{author}{\bibfnamefont{W.~H.} \bibnamefont{Greene}},
  \emph{\bibinfo{title}{Econometric Analysis}} (\bibinfo{publisher}{Pearson,
  8th Edition}, \bibinfo{year}{2017}).

\bibitem[{\citenamefont{Wellman and Way}(2017)}]{Wellman2017}
\bibinfo{author}{\bibfnamefont{M.~P.} \bibnamefont{Wellman}} \bibnamefont{and}
  \bibinfo{author}{\bibfnamefont{E.}~\bibnamefont{Way}}, \bibinfo{journal}{The
  Russell Sage Foundation Journal of the Social Sciences}
  \textbf{\bibinfo{volume}{3(1)}}, \bibinfo{pages}{104} (\bibinfo{year}{2017}).

\bibitem[{\citenamefont{Huang et~al.}(2015)\citenamefont{Huang, Lehalle, and
  Rosenbaum}}]{Huang2015}
\bibinfo{author}{\bibfnamefont{W.}~\bibnamefont{Huang}},
  \bibinfo{author}{\bibfnamefont{C.-A.} \bibnamefont{Lehalle}},
  \bibnamefont{and}
  \bibinfo{author}{\bibfnamefont{M.}~\bibnamefont{Rosenbaum}},
  \bibinfo{journal}{Journal of the American Statistical Association}
  \textbf{\bibinfo{volume}{110}}, \bibinfo{pages}{509} (\bibinfo{year}{2015}).

\bibitem[{\citenamefont{Biondo}(2019)}]{Biondo2019}
\bibinfo{author}{\bibfnamefont{A.~E.} \bibnamefont{Biondo}},
  \bibinfo{journal}{Journal of Economic Interaction and Coordination}
  \textbf{\bibinfo{volume}{14(3)}} (\bibinfo{year}{2019}).

\bibitem[{\citenamefont{Sirignano and Cont}(2019)}]{Sirignano2019}
\bibinfo{author}{\bibfnamefont{J.}~\bibnamefont{Sirignano}} \bibnamefont{and}
  \bibinfo{author}{\bibfnamefont{R.}~\bibnamefont{Cont}},
  \bibinfo{journal}{Quantitative Finance} \textbf{\bibinfo{volume}{19(9)}}
  (\bibinfo{year}{2019}).

\bibitem[{\citenamefont{Sbordone et~al.}(2010)\citenamefont{Sbordone,
  Tambalotti, Rao, and Walsh}}]{Sbordone2010}
\bibinfo{author}{\bibfnamefont{A.~M.} \bibnamefont{Sbordone}},
  \bibinfo{author}{\bibfnamefont{A.}~\bibnamefont{Tambalotti}},
  \bibinfo{author}{\bibfnamefont{K.}~\bibnamefont{Rao}}, \bibnamefont{and}
  \bibinfo{author}{\bibfnamefont{K.~J.} \bibnamefont{Walsh}},
  \bibinfo{journal}{Economic policy review} \textbf{\bibinfo{volume}{16(2)}}
  (\bibinfo{year}{2010}).

\bibitem[{\citenamefont{Wah and Wellman}(2013)}]{Wah2013}
\bibinfo{author}{\bibfnamefont{E.}~\bibnamefont{Wah}} \bibnamefont{and}
  \bibinfo{author}{\bibfnamefont{M.~P.} \bibnamefont{Wellman}},
  \bibinfo{journal}{Proceedings of the fourteenth ACM conference on Electronic
  commerce} pp. \bibinfo{pages}{855--872} (\bibinfo{year}{2013}).

\bibitem[{\citenamefont{Aloud}(2014)}]{Aloud2014}
\bibinfo{author}{\bibfnamefont{M.}~\bibnamefont{Aloud}},
  \bibinfo{journal}{Proceedings in Finance and Risk Perspectives ‘14}
  (\bibinfo{year}{2014}).

\bibitem[{\citenamefont{Benzaquen and Bouchaud}(2018)}]{Benzaquen2018}
\bibinfo{author}{\bibfnamefont{M.}~\bibnamefont{Benzaquen}} \bibnamefont{and}
  \bibinfo{author}{\bibfnamefont{J.-P.} \bibnamefont{Bouchaud}},
  \bibinfo{journal}{The European Physical Journal B}
  \textbf{\bibinfo{volume}{91(23)}} (\bibinfo{year}{2018}).

\bibitem[{\citenamefont{Erev and E.Roth}(2014)}]{ErevRoth2014}
\bibinfo{author}{\bibfnamefont{I.}~\bibnamefont{Erev}} \bibnamefont{and}
  \bibinfo{author}{\bibfnamefont{A.}~\bibnamefont{E.Roth}},
  \bibinfo{journal}{PNAS} \textbf{\bibinfo{volume}{111}},
  \bibinfo{pages}{10818} (\bibinfo{year}{2014}).

\bibitem[{\citenamefont{Martino and Marsili}(2006)}]{Demartino2006}
\bibinfo{author}{\bibfnamefont{A.~D.} \bibnamefont{Martino}} \bibnamefont{and}
  \bibinfo{author}{\bibfnamefont{M.}~\bibnamefont{Marsili}},
  \bibinfo{journal}{Journal of Physics A} \textbf{\bibinfo{volume}{39}},
  \bibinfo{pages}{465} (\bibinfo{year}{2006}).

\bibitem[{\citenamefont{Boero et~al.}(2015)\citenamefont{Boero, Morini,
  Sonnessa, and Terna}}]{Boero2015}
\bibinfo{author}{\bibfnamefont{R.}~\bibnamefont{Boero}},
  \bibinfo{author}{\bibfnamefont{M.}~\bibnamefont{Morini}},
  \bibinfo{author}{\bibfnamefont{M.}~\bibnamefont{Sonnessa}}, \bibnamefont{and}
  \bibinfo{author}{\bibfnamefont{P.}~\bibnamefont{Terna}},
  \emph{\bibinfo{title}{Agent-based models of the economy, from theories to
  applications}} (\bibinfo{publisher}{Palgrave Macmillan},
  \bibinfo{year}{2015}).

\bibitem[{\citenamefont{Gualdi et~al.}(2015)\citenamefont{Gualdi, Tarzia,
  Zamponi, and Bouchaud}}]{Gualdi2015}
\bibinfo{author}{\bibfnamefont{S.}~\bibnamefont{Gualdi}},
  \bibinfo{author}{\bibfnamefont{M.}~\bibnamefont{Tarzia}},
  \bibinfo{author}{\bibfnamefont{F.}~\bibnamefont{Zamponi}}, \bibnamefont{and}
  \bibinfo{author}{\bibfnamefont{J.-P.} \bibnamefont{Bouchaud}},
  \bibinfo{journal}{Journal of Economic Dynamics and Control}
  \textbf{\bibinfo{volume}{50}}, \bibinfo{pages}{29} (\bibinfo{year}{2015}).

\bibitem[{\citenamefont{Eickhoff et~al.}(2018)\citenamefont{Eickhoff, Yeo, and
  Genon}}]{Eickhoff2018}
\bibinfo{author}{\bibfnamefont{S.~B.} \bibnamefont{Eickhoff}},
  \bibinfo{author}{\bibfnamefont{B.~T.~T.} \bibnamefont{Yeo}},
  \bibnamefont{and} \bibinfo{author}{\bibfnamefont{S.}~\bibnamefont{Genon}},
  \bibinfo{journal}{Nature Reviews Neuroscience} \textbf{\bibinfo{volume}{19}},
  \bibinfo{pages}{672} (\bibinfo{year}{2018}).

\bibitem[{\citenamefont{Konovalov and Krajbich}(2016)}]{Konovalov2016}
\bibinfo{author}{\bibfnamefont{A.}~\bibnamefont{Konovalov}} \bibnamefont{and}
  \bibinfo{author}{\bibfnamefont{I.}~\bibnamefont{Krajbich}},
  \bibinfo{journal}{Nature communications} \textbf{\bibinfo{volume}{7}},
  \bibinfo{pages}{12438} (\bibinfo{year}{2016}).

\bibitem[{\citenamefont{Silver et~al.}(2018{\natexlab{a}})\citenamefont{Silver,
  Hubert, Schrittwieser, Antonoglou, Lai, Guez, Lanctot, Sifre, Kumaran,
  Graepel et~al.}}]{Silver2018}
\bibinfo{author}{\bibfnamefont{D.}~\bibnamefont{Silver}},
  \bibinfo{author}{\bibfnamefont{T.}~\bibnamefont{Hubert}},
  \bibinfo{author}{\bibfnamefont{J.}~\bibnamefont{Schrittwieser}},
  \bibinfo{author}{\bibfnamefont{I.}~\bibnamefont{Antonoglou}},
  \bibinfo{author}{\bibfnamefont{M.}~\bibnamefont{Lai}},
  \bibinfo{author}{\bibfnamefont{A.}~\bibnamefont{Guez}},
  \bibinfo{author}{\bibfnamefont{M.}~\bibnamefont{Lanctot}},
  \bibinfo{author}{\bibfnamefont{L.}~\bibnamefont{Sifre}},
  \bibinfo{author}{\bibfnamefont{D.}~\bibnamefont{Kumaran}},
  \bibinfo{author}{\bibfnamefont{T.}~\bibnamefont{Graepel}},
  \bibnamefont{et~al.}, \bibinfo{journal}{Science}
  \textbf{\bibinfo{volume}{362}}, \bibinfo{pages}{1140}
  (\bibinfo{year}{2018}{\natexlab{a}}), ISSN \bibinfo{issn}{0036-8075}.

\bibitem[{\citenamefont{Silver et~al.}(2018{\natexlab{b}})\citenamefont{Silver,
  Schrittwieser, Simonyan, Antonoglou, Huang, Guez, Hubert, Baker, Lai, Bolton
  et~al.}}]{Silver2017}
\bibinfo{author}{\bibfnamefont{D.}~\bibnamefont{Silver}},
  \bibinfo{author}{\bibfnamefont{J.}~\bibnamefont{Schrittwieser}},
  \bibinfo{author}{\bibfnamefont{K.}~\bibnamefont{Simonyan}},
  \bibinfo{author}{\bibfnamefont{I.}~\bibnamefont{Antonoglou}},
  \bibinfo{author}{\bibfnamefont{A.}~\bibnamefont{Huang}},
  \bibinfo{author}{\bibfnamefont{A.}~\bibnamefont{Guez}},
  \bibinfo{author}{\bibfnamefont{T.}~\bibnamefont{Hubert}},
  \bibinfo{author}{\bibfnamefont{L.}~\bibnamefont{Baker}},
  \bibinfo{author}{\bibfnamefont{M.}~\bibnamefont{Lai}},
  \bibinfo{author}{\bibfnamefont{A.}~\bibnamefont{Bolton}},
  \bibnamefont{et~al.}, \bibinfo{journal}{Nature}
  \textbf{\bibinfo{volume}{550}}, \bibinfo{pages}{354}
  (\bibinfo{year}{2018}{\natexlab{b}}).

\bibitem[{\citenamefont{Lefebvre et~al.}(2017)\citenamefont{Lefebvre, Lebreton,
  Meyniel, Bourgeois-Gironde, and Palminteri}}]{Lefebvre2017}
\bibinfo{author}{\bibfnamefont{G.}~\bibnamefont{Lefebvre}},
  \bibinfo{author}{\bibfnamefont{M.}~\bibnamefont{Lebreton}},
  \bibinfo{author}{\bibfnamefont{F.}~\bibnamefont{Meyniel}},
  \bibinfo{author}{\bibfnamefont{S.}~\bibnamefont{Bourgeois-Gironde}},
  \bibnamefont{and}
  \bibinfo{author}{\bibfnamefont{S.}~\bibnamefont{Palminteri}},
  \bibinfo{journal}{Nature Human Behaviour} \textbf{\bibinfo{volume}{1(4)}}
  (\bibinfo{year}{2017}).

\bibitem[{\citenamefont{Palminteri et~al.}(2015)\citenamefont{Palminteri,
  Khamassi, Joffily, and Coricelli}}]{Palminteri2015}
\bibinfo{author}{\bibfnamefont{S.}~\bibnamefont{Palminteri}},
  \bibinfo{author}{\bibfnamefont{M.}~\bibnamefont{Khamassi}},
  \bibinfo{author}{\bibfnamefont{M.}~\bibnamefont{Joffily}}, \bibnamefont{and}
  \bibinfo{author}{\bibfnamefont{G.}~\bibnamefont{Coricelli}},
  \bibinfo{journal}{Nature communications} pp. \bibinfo{pages}{1--14}
  (\bibinfo{year}{2015}).

\bibitem[{\citenamefont{Duncan et~al.}(2018)\citenamefont{Duncan, Doll, Daw,
  and Shohamy}}]{Duncan2018}
\bibinfo{author}{\bibfnamefont{K.}~\bibnamefont{Duncan}},
  \bibinfo{author}{\bibfnamefont{B.~B.} \bibnamefont{Doll}},
  \bibinfo{author}{\bibfnamefont{N.~D.} \bibnamefont{Daw}}, \bibnamefont{and}
  \bibinfo{author}{\bibfnamefont{D.}~\bibnamefont{Shohamy}},
  \bibinfo{journal}{Neuron} \textbf{\bibinfo{volume}{98}}, \bibinfo{pages}{645}
  (\bibinfo{year}{2018}).

\bibitem[{\citenamefont{Momennejad et~al.}(2017)\citenamefont{Momennejad,
  Russek, Cheong, Botvinick, Daw, and Gershman}}]{Momennejad2017}
\bibinfo{author}{\bibfnamefont{I.}~\bibnamefont{Momennejad}},
  \bibinfo{author}{\bibfnamefont{E.}~\bibnamefont{Russek}},
  \bibinfo{author}{\bibfnamefont{J.}~\bibnamefont{Cheong}},
  \bibinfo{author}{\bibfnamefont{M.}~\bibnamefont{Botvinick}},
  \bibinfo{author}{\bibfnamefont{N.~D.} \bibnamefont{Daw}}, \bibnamefont{and}
  \bibinfo{author}{\bibfnamefont{S.~J.} \bibnamefont{Gershman}},
  \bibinfo{journal}{Nature Human Behavior} \textbf{\bibinfo{volume}{1}},
  \bibinfo{pages}{680–692} (\bibinfo{year}{2017}).

\bibitem[{\citenamefont{Gode and Sunder}(1993)}]{Gode1993}
\bibinfo{author}{\bibfnamefont{D.}~\bibnamefont{Gode}} \bibnamefont{and}
  \bibinfo{author}{\bibfnamefont{S.}~\bibnamefont{Sunder}},
  \bibinfo{journal}{Journal of Political Economy}
  \textbf{\bibinfo{volume}{101(1)}} (\bibinfo{year}{1993}).

\bibitem[{\citenamefont{Hu and Lin}(2019)}]{Hu2019}
\bibinfo{author}{\bibfnamefont{Y.-J.} \bibnamefont{Hu}} \bibnamefont{and}
  \bibinfo{author}{\bibfnamefont{S.-J.} \bibnamefont{Lin}},
  \bibinfo{journal}{2019 Amity International Conference on Artificial
  Intelligence}  (\bibinfo{year}{2019}).

\bibitem[{\citenamefont{Neuneier}(1997)}]{Neuneier1997}
\bibinfo{author}{\bibfnamefont{R.}~\bibnamefont{Neuneier}},
  \bibinfo{journal}{Proc. of the 10th International Conference on Neural
  Information Processing Systems}  (\bibinfo{year}{1997}).

\bibitem[{\citenamefont{Deng et~al.}(2017)\citenamefont{Deng, Bao, Kong, Ren,
  and Dai}}]{Deng2017}
\bibinfo{author}{\bibfnamefont{Y.}~\bibnamefont{Deng}},
  \bibinfo{author}{\bibfnamefont{F.}~\bibnamefont{Bao}},
  \bibinfo{author}{\bibfnamefont{Y.}~\bibnamefont{Kong}},
  \bibinfo{author}{\bibfnamefont{Z.}~\bibnamefont{Ren}}, \bibnamefont{and}
  \bibinfo{author}{\bibfnamefont{Q.}~\bibnamefont{Dai}}, \bibinfo{journal}{IEEE
  Trans. on Neural Networks and Learning Systems}
  \textbf{\bibinfo{volume}{28(3)}} (\bibinfo{year}{2017}).

\bibitem[{\citenamefont{Spooner et~al.}(2018)\citenamefont{Spooner, Fearnley,
  Savani, and Koukorinis}}]{Spooner2018}
\bibinfo{author}{\bibfnamefont{T.}~\bibnamefont{Spooner}},
  \bibinfo{author}{\bibfnamefont{J.}~\bibnamefont{Fearnley}},
  \bibinfo{author}{\bibfnamefont{R.}~\bibnamefont{Savani}}, \bibnamefont{and}
  \bibinfo{author}{\bibfnamefont{A.}~\bibnamefont{Koukorinis}},
  \bibinfo{journal}{Proceedings of the 17th AAMAS}  (\bibinfo{year}{2018}).

\bibitem[{\citenamefont{Kim and Markowitz}(1989)}]{Kim1989}
\bibinfo{author}{\bibfnamefont{G.}~\bibnamefont{Kim}} \bibnamefont{and}
  \bibinfo{author}{\bibfnamefont{H.~M.} \bibnamefont{Markowitz}},
  \bibinfo{journal}{Journal of Portfolio Management}
  \textbf{\bibinfo{volume}{16}}, \bibinfo{pages}{45} (\bibinfo{year}{1989}).

\bibitem[{\citenamefont{Levy and Solomon}(1996{\natexlab{a}})}]{Levy1996}
\bibinfo{author}{\bibfnamefont{M.}~\bibnamefont{Levy}} \bibnamefont{and}
  \bibinfo{author}{\bibfnamefont{S.}~\bibnamefont{Solomon}},
  \bibinfo{journal}{International Journal of Modern Physics C}
  \textbf{\bibinfo{volume}{7}}, \bibinfo{pages}{595}
  (\bibinfo{year}{1996}{\natexlab{a}}).

\bibitem[{\citenamefont{Levy et~al.}(1994)\citenamefont{Levy, Levy, and
  Solomon}}]{Levy1994}
\bibinfo{author}{\bibfnamefont{M.}~\bibnamefont{Levy}},
  \bibinfo{author}{\bibfnamefont{H.}~\bibnamefont{Levy}}, \bibnamefont{and}
  \bibinfo{author}{\bibfnamefont{S.}~\bibnamefont{Solomon}},
  \bibinfo{journal}{Economics Letters} \textbf{\bibinfo{volume}{45}},
  \bibinfo{pages}{103} (\bibinfo{year}{1994}).

\bibitem[{\citenamefont{Levy et~al.}(1995)\citenamefont{Levy, Levy, and
  Solomon}}]{Levy1995}
\bibinfo{author}{\bibfnamefont{M.}~\bibnamefont{Levy}},
  \bibinfo{author}{\bibfnamefont{H.}~\bibnamefont{Levy}}, \bibnamefont{and}
  \bibinfo{author}{\bibfnamefont{S.}~\bibnamefont{Solomon}},
  \bibinfo{journal}{Journal de Physique I} \textbf{\bibinfo{volume}{5}},
  \bibinfo{pages}{1087} (\bibinfo{year}{1995}).

\bibitem[{\citenamefont{Levy and Solomon}(1996{\natexlab{b}})}]{Levy1996b}
\bibinfo{author}{\bibfnamefont{M.}~\bibnamefont{Levy}} \bibnamefont{and}
  \bibinfo{author}{\bibfnamefont{S.}~\bibnamefont{Solomon}},
  \bibinfo{journal}{International Journal of Modern Physics C}
  \textbf{\bibinfo{volume}{7}}, \bibinfo{pages}{65}
  (\bibinfo{year}{1996}{\natexlab{b}}).

\bibitem[{\citenamefont{Levy et~al.}(1996)\citenamefont{Levy, Persky, and
  Solomon}}]{Levy1996c}
\bibinfo{author}{\bibfnamefont{M.}~\bibnamefont{Levy}},
  \bibinfo{author}{\bibfnamefont{N.}~\bibnamefont{Persky}}, \bibnamefont{and}
  \bibinfo{author}{\bibfnamefont{S.}~\bibnamefont{Solomon}},
  \bibinfo{journal}{International Journal of High Speed Computing}
  \textbf{\bibinfo{volume}{8}}, \bibinfo{pages}{93} (\bibinfo{year}{1996}).

\bibitem[{\citenamefont{Levy et~al.}(1997)\citenamefont{Levy, Levy, and
  Solomon}}]{Levy1997}
\bibinfo{author}{\bibfnamefont{M.}~\bibnamefont{Levy}},
  \bibinfo{author}{\bibfnamefont{H.}~\bibnamefont{Levy}}, \bibnamefont{and}
  \bibinfo{author}{\bibfnamefont{S.}~\bibnamefont{Solomon}},
  \bibinfo{journal}{Physica A} \textbf{\bibinfo{volume}{242}},
  \bibinfo{pages}{90} (\bibinfo{year}{1997}).

\bibitem[{\citenamefont{Levy et~al.}(2000)\citenamefont{Levy, Levy, and
  Solomon}}]{Levy2000}
\bibinfo{author}{\bibfnamefont{M.}~\bibnamefont{Levy}},
  \bibinfo{author}{\bibfnamefont{H.}~\bibnamefont{Levy}}, \bibnamefont{and}
  \bibinfo{author}{\bibfnamefont{S.}~\bibnamefont{Solomon}},
  \emph{\bibinfo{title}{Microscopic Simulation of Financial Markets}}
  (\bibinfo{publisher}{Academic Press, New York}, \bibinfo{year}{2000}).

\bibitem[{\citenamefont{Cont and Bouchaud}(2000)}]{Cont2000}
\bibinfo{author}{\bibfnamefont{R.}~\bibnamefont{Cont}} \bibnamefont{and}
  \bibinfo{author}{\bibfnamefont{J.~P.} \bibnamefont{Bouchaud}},
  \bibinfo{journal}{Macroeconomic Dynamics} \textbf{\bibinfo{volume}{4}},
  \bibinfo{pages}{170} (\bibinfo{year}{2000}).

\bibitem[{\citenamefont{Solomon et~al.}(2000)\citenamefont{Solomon, Weisbuch,
  de~Arcangelis, Jan, and Stauffer}}]{Solomon2000}
\bibinfo{author}{\bibfnamefont{S.}~\bibnamefont{Solomon}},
  \bibinfo{author}{\bibfnamefont{G.}~\bibnamefont{Weisbuch}},
  \bibinfo{author}{\bibfnamefont{L.}~\bibnamefont{de~Arcangelis}},
  \bibinfo{author}{\bibfnamefont{N.}~\bibnamefont{Jan}}, \bibnamefont{and}
  \bibinfo{author}{\bibfnamefont{D.}~\bibnamefont{Stauffer}},
  \bibinfo{journal}{Physica A} \textbf{\bibinfo{volume}{277(1)}},
  \bibinfo{pages}{239} (\bibinfo{year}{2000}).

\bibitem[{\citenamefont{Lux and Marchesi}(1999)}]{Lux1999}
\bibinfo{author}{\bibfnamefont{T.}~\bibnamefont{Lux}} \bibnamefont{and}
  \bibinfo{author}{\bibfnamefont{M.}~\bibnamefont{Marchesi}},
  \bibinfo{journal}{Nature} \textbf{\bibinfo{volume}{397}},
  \bibinfo{pages}{498} (\bibinfo{year}{1999}).

\bibitem[{\citenamefont{Lux and Marchesi}(2000)}]{Lux2000}
\bibinfo{author}{\bibfnamefont{T.}~\bibnamefont{Lux}} \bibnamefont{and}
  \bibinfo{author}{\bibfnamefont{M.}~\bibnamefont{Marchesi}},
  \bibinfo{journal}{Journal of Theoretical and Applied Finance}
  \textbf{\bibinfo{volume}{3}}, \bibinfo{pages}{67} (\bibinfo{year}{2000}).

\bibitem[{\citenamefont{Donangelo et~al.}(2000)\citenamefont{Donangelo, Hansen,
  Sneppen, and Souza}}]{Donangelo2000}
\bibinfo{author}{\bibfnamefont{R.}~\bibnamefont{Donangelo}},
  \bibinfo{author}{\bibfnamefont{A.}~\bibnamefont{Hansen}},
  \bibinfo{author}{\bibfnamefont{K.}~\bibnamefont{Sneppen}}, \bibnamefont{and}
  \bibinfo{author}{\bibfnamefont{S.~R.} \bibnamefont{Souza}},
  \bibinfo{journal}{Physica A} \textbf{\bibinfo{volume}{283}},
  \bibinfo{pages}{469} (\bibinfo{year}{2000}).

\bibitem[{\citenamefont{Donangelo and Sneppen}(2000)}]{Donangelo2000b}
\bibinfo{author}{\bibfnamefont{R.}~\bibnamefont{Donangelo}} \bibnamefont{and}
  \bibinfo{author}{\bibfnamefont{K.}~\bibnamefont{Sneppen}},
  \bibinfo{journal}{Physica A} \textbf{\bibinfo{volume}{276}},
  \bibinfo{pages}{572} (\bibinfo{year}{2000}).

\bibitem[{\citenamefont{Bak et~al.}(1999)\citenamefont{Bak, Norrelykke, and
  Shubik}}]{Bak1999}
\bibinfo{author}{\bibfnamefont{P.}~\bibnamefont{Bak}},
  \bibinfo{author}{\bibfnamefont{S.}~\bibnamefont{Norrelykke}},
  \bibnamefont{and} \bibinfo{author}{\bibfnamefont{M.}~\bibnamefont{Shubik}},
  \bibinfo{journal}{Physical Review E} \textbf{\bibinfo{volume}{60}},
  \bibinfo{pages}{2528} (\bibinfo{year}{1999}).

\bibitem[{\citenamefont{Bak et~al.}(2001)\citenamefont{Bak, Norrelykke, and
  Shubik}}]{Bak2001}
\bibinfo{author}{\bibfnamefont{P.}~\bibnamefont{Bak}},
  \bibinfo{author}{\bibfnamefont{S.}~\bibnamefont{Norrelykke}},
  \bibnamefont{and} \bibinfo{author}{\bibfnamefont{M.}~\bibnamefont{Shubik}},
  \bibinfo{journal}{Quantitative Finance} \textbf{\bibinfo{volume}{1}},
  \bibinfo{pages}{186} (\bibinfo{year}{2001}).

\bibitem[{\citenamefont{Huang and Solomon}(2000)}]{Huang2000}
\bibinfo{author}{\bibfnamefont{Z.~F.} \bibnamefont{Huang}} \bibnamefont{and}
  \bibinfo{author}{\bibfnamefont{S.}~\bibnamefont{Solomon}},
  \bibinfo{journal}{European Physical Journal B} \textbf{\bibinfo{volume}{20}},
  \bibinfo{pages}{601} (\bibinfo{year}{2000}).

\bibitem[{\citenamefont{Lipski and Kutner}(2013)}]{Lipski2013}
\bibinfo{author}{\bibfnamefont{J.}~\bibnamefont{Lipski}} \bibnamefont{and}
  \bibinfo{author}{\bibfnamefont{R.}~\bibnamefont{Kutner}},
  \bibinfo{journal}{arXiv:1310.0762}  (\bibinfo{year}{2013}).

\bibitem[{\citenamefont{Barde}(2015)}]{Barde2015}
\bibinfo{author}{\bibfnamefont{S.}~\bibnamefont{Barde}},
  \bibinfo{journal}{University of Kent, School of Economics Discussion Papers}
  \textbf{\bibinfo{volume}{04}} (\bibinfo{year}{2015}).

\bibitem[{\citenamefont{Bouchaud}(2018)}]{Bouchaud2018}
\bibinfo{author}{\bibfnamefont{J.-P.} \bibnamefont{Bouchaud}},
  \bibinfo{journal}{Handbook of Computational Economics}
  \textbf{\bibinfo{volume}{4}} (\bibinfo{year}{2018}).

\bibitem[{\citenamefont{Dodonova and Khoroshilov}(2018)}]{Dodonova2018}
\bibinfo{author}{\bibfnamefont{A.}~\bibnamefont{Dodonova}} \bibnamefont{and}
  \bibinfo{author}{\bibfnamefont{Y.}~\bibnamefont{Khoroshilov}},
  \bibinfo{journal}{Manag Decis Econ} \textbf{\bibinfo{volume}{39}}
  (\bibinfo{year}{2018}).

\bibitem[{\citenamefont{Naik et~al.}(2018)\citenamefont{Naik, Gupta, and
  Padhi}}]{Naik2018}
\bibinfo{author}{\bibfnamefont{P.~K.} \bibnamefont{Naik}},
  \bibinfo{author}{\bibfnamefont{R.}~\bibnamefont{Gupta}}, \bibnamefont{and}
  \bibinfo{author}{\bibfnamefont{P.}~\bibnamefont{Padhi}}, \bibinfo{journal}{J
  Dev Areas} \textbf{\bibinfo{volume}{52(1)}} (\bibinfo{year}{2018}).

\bibitem[{\citenamefont{Cont}(2005)}]{Cont2005}
\bibinfo{author}{\bibfnamefont{R.}~\bibnamefont{Cont}},
  \emph{\bibinfo{title}{Volatility Clustering in Financial Markets: Empirical
  Facts and Agent-Based Models}} (\bibinfo{publisher}{A Kirman and G Teyssiere:
  Long memory in economics, Springer}, \bibinfo{year}{2005}).

\bibitem[{\citenamefont{Cristelli}(2014)}]{Cristelli2014}
\bibinfo{author}{\bibfnamefont{M.}~\bibnamefont{Cristelli}},
  \emph{\bibinfo{title}{Complexity in Financial Markets}}
  (\bibinfo{publisher}{Springer}, \bibinfo{year}{2014}).

\bibitem[{\citenamefont{Fama}(1970)}]{Fama1970}
\bibinfo{author}{\bibfnamefont{E.}~\bibnamefont{Fama}},
  \bibinfo{journal}{Journal of Finance} \textbf{\bibinfo{volume}{25}},
  \bibinfo{pages}{383} (\bibinfo{year}{1970}).

\bibitem[{\citenamefont{Bera et~al.}(2015)\citenamefont{Bera, Ivliev, and
  Lillo}}]{Bera2015}
\bibinfo{author}{\bibfnamefont{A.~K.} \bibnamefont{Bera}},
  \bibinfo{author}{\bibfnamefont{S.}~\bibnamefont{Ivliev}}, \bibnamefont{and}
  \bibinfo{author}{\bibfnamefont{F.}~\bibnamefont{Lillo}},
  \emph{\bibinfo{title}{Financial Econometrics and Empirical Market
  Microstructure}} (\bibinfo{publisher}{Springer}, \bibinfo{year}{2015}).

\bibitem[{\citenamefont{Potters and Bouchaud}(2001)}]{Potters2001}
\bibinfo{author}{\bibfnamefont{M.}~\bibnamefont{Potters}} \bibnamefont{and}
  \bibinfo{author}{\bibfnamefont{J.-P.} \bibnamefont{Bouchaud}},
  \bibinfo{journal}{Physica A} \textbf{\bibinfo{volume}{299}},
  \bibinfo{pages}{60} (\bibinfo{year}{2001}).

\bibitem[{\citenamefont{Plerou et~al.}(1999)\citenamefont{Plerou, Gopikrishnan,
  Amaral, Meyer, and Stanley}}]{Plerou1999}
\bibinfo{author}{\bibfnamefont{V.}~\bibnamefont{Plerou}},
  \bibinfo{author}{\bibfnamefont{P.}~\bibnamefont{Gopikrishnan}},
  \bibinfo{author}{\bibfnamefont{L.~A.} \bibnamefont{Amaral}},
  \bibinfo{author}{\bibfnamefont{M.}~\bibnamefont{Meyer}}, \bibnamefont{and}
  \bibinfo{author}{\bibfnamefont{H.~E.} \bibnamefont{Stanley}},
  \bibinfo{journal}{Physical Review E} \textbf{\bibinfo{volume}{60(6)}},
  \bibinfo{pages}{6519} (\bibinfo{year}{1999}).

\bibitem[{\citenamefont{Weron}(2001)}]{Weron2001}
\bibinfo{author}{\bibfnamefont{R.}~\bibnamefont{Weron}},
  \bibinfo{journal}{International Journal of Modern Physics C}
  \textbf{\bibinfo{volume}{12}}, \bibinfo{pages}{209} (\bibinfo{year}{2001}).

\bibitem[{\citenamefont{Eisler and Kertesz}(2006)}]{Eisler2006}
\bibinfo{author}{\bibfnamefont{Z.}~\bibnamefont{Eisler}} \bibnamefont{and}
  \bibinfo{author}{\bibfnamefont{J.}~\bibnamefont{Kertesz}},
  \bibinfo{journal}{European Physical Journal B} \textbf{\bibinfo{volume}{51}},
  \bibinfo{pages}{145} (\bibinfo{year}{2006}).

\bibitem[{\citenamefont{Mandelbrot}(1963)}]{Mandelbrot1963}
\bibinfo{author}{\bibfnamefont{B.}~\bibnamefont{Mandelbrot}},
  \bibinfo{journal}{The Journal of Business} pp. \bibinfo{pages}{394--419}
  (\bibinfo{year}{1963}).

\bibitem[{\citenamefont{Cont}(2001)}]{Cont2001}
\bibinfo{author}{\bibfnamefont{R.}~\bibnamefont{Cont}},
  \bibinfo{journal}{Quantitative Finance} \textbf{\bibinfo{volume}{1}},
  \bibinfo{pages}{223} (\bibinfo{year}{2001}).

\bibitem[{\citenamefont{Bouchaud et~al.}(1997)\citenamefont{Bouchaud, Cont, and
  Potters}}]{Bouchaud1997}
\bibinfo{author}{\bibfnamefont{J.}~\bibnamefont{Bouchaud}},
  \bibinfo{author}{\bibfnamefont{R.}~\bibnamefont{Cont}}, \bibnamefont{and}
  \bibinfo{author}{\bibfnamefont{M.}~\bibnamefont{Potters}},
  \emph{\bibinfo{title}{Scale Invariance and Beyond, Proc. CNRS Workshop on
  Scale Invariance, Les Houches}} (\bibinfo{publisher}{Springer},
  \bibinfo{year}{1997}).

\bibitem[{\citenamefont{Ding et~al.}(1993)\citenamefont{Ding, Engle, and
  Granger}}]{Ding1993}
\bibinfo{author}{\bibfnamefont{Z.}~\bibnamefont{Ding}},
  \bibinfo{author}{\bibfnamefont{R.}~\bibnamefont{Engle}}, \bibnamefont{and}
  \bibinfo{author}{\bibfnamefont{C.}~\bibnamefont{Granger}},
  \bibinfo{journal}{Journal of Empirical Finance} \textbf{\bibinfo{volume}{1}},
  \bibinfo{pages}{83} (\bibinfo{year}{1993}).

\bibitem[{\citenamefont{Lobato and Savin}(1998)}]{Lobato1998}
\bibinfo{author}{\bibfnamefont{I.~N.} \bibnamefont{Lobato}} \bibnamefont{and}
  \bibinfo{author}{\bibfnamefont{N.~E.} \bibnamefont{Savin}},
  \bibinfo{journal}{Journal of Business and Economics Statistics}
  \textbf{\bibinfo{volume}{16}}, \bibinfo{pages}{261} (\bibinfo{year}{1998}).

\bibitem[{\citenamefont{Vandewalle and Ausloos}(1997)}]{Vandewalle1997}
\bibinfo{author}{\bibfnamefont{N.}~\bibnamefont{Vandewalle}} \bibnamefont{and}
  \bibinfo{author}{\bibfnamefont{M.}~\bibnamefont{Ausloos}},
  \bibinfo{journal}{Physica A} \textbf{\bibinfo{volume}{246}},
  \bibinfo{pages}{454} (\bibinfo{year}{1997}).

\bibitem[{\citenamefont{Mandelbrot et~al.}(1997)\citenamefont{Mandelbrot,
  Fisher, and Calvet}}]{Mandelbrot1997}
\bibinfo{author}{\bibfnamefont{B.}~\bibnamefont{Mandelbrot}},
  \bibinfo{author}{\bibfnamefont{A.}~\bibnamefont{Fisher}}, \bibnamefont{and}
  \bibinfo{author}{\bibfnamefont{L.}~\bibnamefont{Calvet}},
  \emph{\bibinfo{title}{A multifractal model of asset returns}}
  (\bibinfo{publisher}{Cowles Foundation for Research and Economics},
  \bibinfo{year}{1997}).

\bibitem[{\citenamefont{Engle}(1982)}]{Engle1982}
\bibinfo{author}{\bibfnamefont{R.~F.} \bibnamefont{Engle}},
  \bibinfo{journal}{Econometrica} \textbf{\bibinfo{volume}{50(4)}},
  \bibinfo{pages}{987} (\bibinfo{year}{1982}).

\bibitem[{\citenamefont{de~Vries and Leuven}(1994)}]{Devries1994}
\bibinfo{author}{\bibfnamefont{C.}~\bibnamefont{de~Vries}} \bibnamefont{and}
  \bibinfo{author}{\bibfnamefont{K.}~\bibnamefont{Leuven}},
  \emph{\bibinfo{title}{Stylized facts of nominal exchange rate returns}}
  (\bibinfo{publisher}{Working Papers from Purdue University, Krannert School
  of Management - Center for International Business Education and Research
  (CIBER)}, \bibinfo{year}{1994}).

\bibitem[{\citenamefont{Pagan}(1996)}]{Pagan1996}
\bibinfo{author}{\bibfnamefont{A.}~\bibnamefont{Pagan}},
  \bibinfo{journal}{Journal of Empirical Finance} \textbf{\bibinfo{volume}{3}},
  \bibinfo{pages}{15} (\bibinfo{year}{1996}).

\bibitem[{\citenamefont{Sutton and Barto}(1998)}]{SuttonBarto}
\bibinfo{author}{\bibfnamefont{R.}~\bibnamefont{Sutton}} \bibnamefont{and}
  \bibinfo{author}{\bibfnamefont{A.}~\bibnamefont{Barto}},
  \emph{\bibinfo{title}{Reinforcement Learning: An Introduction}}
  (\bibinfo{publisher}{MIT Press Cambridge MA}, \bibinfo{year}{1998}).

\bibitem[{\citenamefont{Wiering and van Otterlo}(2012)}]{Wiering2012}
\bibinfo{author}{\bibfnamefont{M.}~\bibnamefont{Wiering}} \bibnamefont{and}
  \bibinfo{author}{\bibfnamefont{M.}~\bibnamefont{van Otterlo}},
  \emph{\bibinfo{title}{Reinforcement Learning: State-of-the-Art}}
  (\bibinfo{publisher}{Springer, Berlin, Heidelberg}, \bibinfo{year}{2012}).

\bibitem[{\citenamefont{Szepesvari}(2010)}]{Csaba2010}
\bibinfo{author}{\bibfnamefont{C.}~\bibnamefont{Szepesvari}},
  \emph{\bibinfo{title}{Algorithms for Reinforcement Learning}}
  (\bibinfo{publisher}{Morgan and Claypool Publishers}, \bibinfo{year}{2010}).

\bibitem[{\citenamefont{Sutton et~al.}(2000)\citenamefont{Sutton, McAllester,
  Singh, and Mansour}}]{Sutton2000}
\bibinfo{author}{\bibfnamefont{R.~S.} \bibnamefont{Sutton}},
  \bibinfo{author}{\bibfnamefont{D.}~\bibnamefont{McAllester}},
  \bibinfo{author}{\bibfnamefont{S.}~\bibnamefont{Singh}}, \bibnamefont{and}
  \bibinfo{author}{\bibfnamefont{Y.}~\bibnamefont{Mansour}},
  \bibinfo{journal}{Advances in Neural Information Processing Systems}
  \textbf{\bibinfo{volume}{12}}, \bibinfo{pages}{1057} (\bibinfo{year}{2000}).

\bibitem[{\citenamefont{Silver et~al.}(2014)\citenamefont{Silver, Lever, Heess,
  Degris, Wierstra, and Riedmiller}}]{Silver2014}
\bibinfo{author}{\bibfnamefont{D.}~\bibnamefont{Silver}},
  \bibinfo{author}{\bibfnamefont{G.}~\bibnamefont{Lever}},
  \bibinfo{author}{\bibfnamefont{N.}~\bibnamefont{Heess}},
  \bibinfo{author}{\bibfnamefont{T.}~\bibnamefont{Degris}},
  \bibinfo{author}{\bibfnamefont{D.}~\bibnamefont{Wierstra}}, \bibnamefont{and}
  \bibinfo{author}{\bibfnamefont{M.}~\bibnamefont{Riedmiller}},
  \bibinfo{journal}{Proceedings of the 31st International Conference on Machine
  Learning} \textbf{\bibinfo{volume}{32}} (\bibinfo{year}{2014}).

\bibitem[{\citenamefont{Grondman et~al.}(2012)\citenamefont{Grondman, Busoniu,
  Lopes, and Babuska}}]{Grondman2012}
\bibinfo{author}{\bibfnamefont{I.}~\bibnamefont{Grondman}},
  \bibinfo{author}{\bibfnamefont{L.}~\bibnamefont{Busoniu}},
  \bibinfo{author}{\bibfnamefont{G.}~\bibnamefont{Lopes}}, \bibnamefont{and}
  \bibinfo{author}{\bibfnamefont{R.}~\bibnamefont{Babuska}},
  \bibinfo{journal}{IEEE Transactions on Systems Man and Cybernetics}
  \textbf{\bibinfo{volume}{42}}, \bibinfo{pages}{1291} (\bibinfo{year}{2012}).

\bibitem[{\citenamefont{Wang et~al.}(2018)\citenamefont{Wang, Kurth-Nelson,
  Kumaran, Tirumala, Soyer, Leibo, Hassabis, and Botvinick}}]{Wang2018}
\bibinfo{author}{\bibfnamefont{J.~X.} \bibnamefont{Wang}},
  \bibinfo{author}{\bibfnamefont{Z.}~\bibnamefont{Kurth-Nelson}},
  \bibinfo{author}{\bibfnamefont{D.}~\bibnamefont{Kumaran}},
  \bibinfo{author}{\bibfnamefont{D.}~\bibnamefont{Tirumala}},
  \bibinfo{author}{\bibfnamefont{H.}~\bibnamefont{Soyer}},
  \bibinfo{author}{\bibfnamefont{J.~Z.} \bibnamefont{Leibo}},
  \bibinfo{author}{\bibfnamefont{D.}~\bibnamefont{Hassabis}}, \bibnamefont{and}
  \bibinfo{author}{\bibfnamefont{M.}~\bibnamefont{Botvinick}},
  \bibinfo{journal}{Nature Neuroscience} \textbf{\bibinfo{volume}{21}},
  \bibinfo{pages}{860} (\bibinfo{year}{2018}).

\bibitem[{\citenamefont{Duan et~al.}(2016)\citenamefont{Duan, Schulman, Chen,
  Bartlett, Sutskever, and Abbeel}}]{Duan2016}
\bibinfo{author}{\bibfnamefont{Y.}~\bibnamefont{Duan}},
  \bibinfo{author}{\bibfnamefont{J.}~\bibnamefont{Schulman}},
  \bibinfo{author}{\bibfnamefont{X.}~\bibnamefont{Chen}},
  \bibinfo{author}{\bibfnamefont{P.~L.} \bibnamefont{Bartlett}},
  \bibinfo{author}{\bibfnamefont{I.}~\bibnamefont{Sutskever}},
  \bibnamefont{and} \bibinfo{author}{\bibfnamefont{P.}~\bibnamefont{Abbeel}},
  \bibinfo{journal}{arXiv:1611.02779}  (\bibinfo{year}{2016}).

\bibitem[{\citenamefont{Heinrich}(2017)}]{Heinrich2017}
\bibinfo{author}{\bibfnamefont{J.}~\bibnamefont{Heinrich}}, Ph.D. thesis,
  \bibinfo{school}{University College London} (\bibinfo{year}{2017}).

\bibitem[{\citenamefont{Mnih et~al.}(2016)\citenamefont{Mnih, Badia, Mirza,
  Graves, Lillicrap, Harley, Silver, and Kavukcuoglu}}]{Mnih2016}
\bibinfo{author}{\bibfnamefont{V.}~\bibnamefont{Mnih}},
  \bibinfo{author}{\bibfnamefont{A.~P.} \bibnamefont{Badia}},
  \bibinfo{author}{\bibfnamefont{M.}~\bibnamefont{Mirza}},
  \bibinfo{author}{\bibfnamefont{A.}~\bibnamefont{Graves}},
  \bibinfo{author}{\bibfnamefont{T.~P.} \bibnamefont{Lillicrap}},
  \bibinfo{author}{\bibfnamefont{T.}~\bibnamefont{Harley}},
  \bibinfo{author}{\bibfnamefont{D.}~\bibnamefont{Silver}}, \bibnamefont{and}
  \bibinfo{author}{\bibfnamefont{K.}~\bibnamefont{Kavukcuoglu}},
  \bibinfo{journal}{arXiv:1602.01783}  (\bibinfo{year}{2016}).

\bibitem[{\citenamefont{Andreas et~al.}(2017)\citenamefont{Andreas, Klein, and
  Levine}}]{Andreas2017}
\bibinfo{author}{\bibfnamefont{J.}~\bibnamefont{Andreas}},
  \bibinfo{author}{\bibfnamefont{D.}~\bibnamefont{Klein}}, \bibnamefont{and}
  \bibinfo{author}{\bibfnamefont{S.}~\bibnamefont{Levine}},
  \bibinfo{journal}{International Conference on Machine Learning}
  (\bibinfo{year}{2017}).

\bibitem[{\citenamefont{Silver et~al.}(2016)\citenamefont{Silver, Huang,
  Maddison, Guez, Sifre, van~den Driessche, Schrittwieser, and
  et~al.}}]{Silver2016}
\bibinfo{author}{\bibfnamefont{D.}~\bibnamefont{Silver}},
  \bibinfo{author}{\bibfnamefont{A.}~\bibnamefont{Huang}},
  \bibinfo{author}{\bibfnamefont{C.~J.} \bibnamefont{Maddison}},
  \bibinfo{author}{\bibfnamefont{A.}~\bibnamefont{Guez}},
  \bibinfo{author}{\bibfnamefont{L.}~\bibnamefont{Sifre}},
  \bibinfo{author}{\bibfnamefont{G.}~\bibnamefont{van~den Driessche}},
  \bibinfo{author}{\bibfnamefont{J.}~\bibnamefont{Schrittwieser}},
  \bibnamefont{and} \bibinfo{author}{\bibnamefont{et~al.}},
  \bibinfo{journal}{Nature} \textbf{\bibinfo{volume}{529}},
  \bibinfo{pages}{484} (\bibinfo{year}{2016}).

\bibitem[{\citenamefont{Tessler et~al.}(2016)\citenamefont{Tessler, Givony,
  Zahavy, Mankowitz, and Mannor}}]{Tessler2016}
\bibinfo{author}{\bibfnamefont{C.}~\bibnamefont{Tessler}},
  \bibinfo{author}{\bibfnamefont{S.}~\bibnamefont{Givony}},
  \bibinfo{author}{\bibfnamefont{T.}~\bibnamefont{Zahavy}},
  \bibinfo{author}{\bibfnamefont{D.~J.} \bibnamefont{Mankowitz}},
  \bibnamefont{and} \bibinfo{author}{\bibfnamefont{S.}~\bibnamefont{Mannor}},
  \bibinfo{journal}{arXiv:1604.07255}  (\bibinfo{year}{2016}).

\bibitem[{\citenamefont{Bhatnagara and Panigrahi}(2006)}]{Bhatnagara2006}
\bibinfo{author}{\bibfnamefont{S.}~\bibnamefont{Bhatnagara}} \bibnamefont{and}
  \bibinfo{author}{\bibfnamefont{J.~R.} \bibnamefont{Panigrahi}},
  \bibinfo{journal}{Automatica} \textbf{\bibinfo{volume}{42}}
  (\bibinfo{year}{2006}).

\bibitem[{\citenamefont{Liang et~al.}(2017)\citenamefont{Liang, Yang, Tu, and
  Xu}}]{Liang2017}
\bibinfo{author}{\bibfnamefont{H.}~\bibnamefont{Liang}},
  \bibinfo{author}{\bibfnamefont{L.}~\bibnamefont{Yang}},
  \bibinfo{author}{\bibfnamefont{H.~C.~W.} \bibnamefont{Tu}}, \bibnamefont{and}
  \bibinfo{author}{\bibfnamefont{M.}~\bibnamefont{Xu}}, \bibinfo{journal}{2017
  Chinese Automation Congress}  (\bibinfo{year}{2017}).

\bibitem[{\citenamefont{Ng et~al.}(1999)\citenamefont{Ng, Harada, and
  Russell}}]{Ng1999}
\bibinfo{author}{\bibfnamefont{A.~Y.} \bibnamefont{Ng}},
  \bibinfo{author}{\bibfnamefont{D.}~\bibnamefont{Harada}}, \bibnamefont{and}
  \bibinfo{author}{\bibfnamefont{S.}~\bibnamefont{Russell}},
  \bibinfo{journal}{International Conference on Machine Learning}
  (\bibinfo{year}{1999}).

\bibitem[{\citenamefont{Abbeel et~al.}(2010)\citenamefont{Abbeel, Coates, and
  Ng}}]{Abbeel2010}
\bibinfo{author}{\bibfnamefont{P.}~\bibnamefont{Abbeel}},
  \bibinfo{author}{\bibfnamefont{A.}~\bibnamefont{Coates}}, \bibnamefont{and}
  \bibinfo{author}{\bibfnamefont{A.~Y.} \bibnamefont{Ng}},
  \bibinfo{journal}{The International Journal of Robotics Research}
  (\bibinfo{year}{2010}).

\bibitem[{\citenamefont{Keramati and Gutkin}(2014)}]{Keramati2014}
\bibinfo{author}{\bibfnamefont{M.}~\bibnamefont{Keramati}} \bibnamefont{and}
  \bibinfo{author}{\bibfnamefont{B.}~\bibnamefont{Gutkin}},
  \bibinfo{journal}{Elife} \textbf{\bibinfo{volume}{3}} (\bibinfo{year}{2014}).

\bibitem[{\citenamefont{Keramati and Gutkin}(2011)}]{Keramati2011}
\bibinfo{author}{\bibfnamefont{M.}~\bibnamefont{Keramati}} \bibnamefont{and}
  \bibinfo{author}{\bibfnamefont{B.}~\bibnamefont{Gutkin}},
  \bibinfo{journal}{NIPS}  (\bibinfo{year}{2011}).

\bibitem[{\citenamefont{Bavard et~al.}(2018)\citenamefont{Bavard, Lebreton,
  Khamassi, Coricelli, and Palminteri}}]{Bavard2018}
\bibinfo{author}{\bibfnamefont{S.}~\bibnamefont{Bavard}},
  \bibinfo{author}{\bibfnamefont{M.}~\bibnamefont{Lebreton}},
  \bibinfo{author}{\bibfnamefont{M.}~\bibnamefont{Khamassi}},
  \bibinfo{author}{\bibfnamefont{G.}~\bibnamefont{Coricelli}},
  \bibnamefont{and}
  \bibinfo{author}{\bibfnamefont{S.}~\bibnamefont{Palminteri}},
  \bibinfo{journal}{Nature Communications} \textbf{\bibinfo{volume}{4503}}
  (\bibinfo{year}{2018}).

\bibitem[{\citenamefont{Watkins and Dayan}(1992)}]{Watkins1992}
\bibinfo{author}{\bibfnamefont{C.~J. C.~H.} \bibnamefont{Watkins}}
  \bibnamefont{and} \bibinfo{author}{\bibfnamefont{P.}~\bibnamefont{Dayan}},
  \bibinfo{journal}{Machine learning} \textbf{\bibinfo{volume}{8(3-4)}},
  \bibinfo{pages}{279} (\bibinfo{year}{1992}).

\bibitem[{\citenamefont{Ross et~al.}(2011)\citenamefont{Ross, Pineau,
  Chaib-draa, and Kreitmann}}]{Ross2011}
\bibinfo{author}{\bibfnamefont{S.}~\bibnamefont{Ross}},
  \bibinfo{author}{\bibfnamefont{J.}~\bibnamefont{Pineau}},
  \bibinfo{author}{\bibfnamefont{B.}~\bibnamefont{Chaib-draa}},
  \bibnamefont{and}
  \bibinfo{author}{\bibfnamefont{P.}~\bibnamefont{Kreitmann}},
  \bibinfo{journal}{Journal of Machine Learning Research}
  \textbf{\bibinfo{volume}{12}}, \bibinfo{pages}{1729} (\bibinfo{year}{2011}).

\bibitem[{\citenamefont{Katt et~al.}(2017)\citenamefont{Katt, Oliehoek, and
  Amato}}]{Katt2017}
\bibinfo{author}{\bibfnamefont{S.}~\bibnamefont{Katt}},
  \bibinfo{author}{\bibfnamefont{F.~A.} \bibnamefont{Oliehoek}},
  \bibnamefont{and} \bibinfo{author}{\bibfnamefont{C.}~\bibnamefont{Amato}},
  \bibinfo{journal}{Proceedings of the 34th International Conference on Machine
  Learning}  (\bibinfo{year}{2017}).

\bibitem[{\citenamefont{Pinto et~al.}(2017)\citenamefont{Pinto, Davidson,
  Sukthankar, and Gupta}}]{Pinto2017}
\bibinfo{author}{\bibfnamefont{L.}~\bibnamefont{Pinto}},
  \bibinfo{author}{\bibfnamefont{J.}~\bibnamefont{Davidson}},
  \bibinfo{author}{\bibfnamefont{R.}~\bibnamefont{Sukthankar}},
  \bibnamefont{and} \bibinfo{author}{\bibfnamefont{A.}~\bibnamefont{Gupta}},
  \bibinfo{journal}{arXiv:1703.02702}  (\bibinfo{year}{2017}).

\bibitem[{Div()}]{DividendYield}
\emph{\bibinfo{title}{Dividend yield for stocks in the dow jones industrial
  average}}, \bibinfo{note}{accessed: 2019-05-17},
  \urlprefix\url{http://indexarb.com/dividendYieldSorteddj.html}.

\bibitem[{\citenamefont{Franke and Westerhoff}(2011)}]{Franke2011}
\bibinfo{author}{\bibfnamefont{R.}~\bibnamefont{Franke}} \bibnamefont{and}
  \bibinfo{author}{\bibfnamefont{F.}~\bibnamefont{Westerhoff}},
  \bibinfo{journal}{BERG Working Paper Series on Government and Growth}
  \textbf{\bibinfo{volume}{78}} (\bibinfo{year}{2011}).

\bibitem[{\citenamefont{Chiarella et~al.}(2007)\citenamefont{Chiarella, Iori,
  and Perell}}]{Chiarella2007}
\bibinfo{author}{\bibfnamefont{C.}~\bibnamefont{Chiarella}},
  \bibinfo{author}{\bibfnamefont{G.}~\bibnamefont{Iori}}, \bibnamefont{and}
  \bibinfo{author}{\bibfnamefont{J.}~\bibnamefont{Perell}},
  \bibinfo{journal}{arXiv:0711.3581}  (\bibinfo{year}{2007}).

\bibitem[{\citenamefont{Vernimmen et~al.}(2014)\citenamefont{Vernimmen, Quiry,
  Dallocchio, Fur, and Salvi}}]{Vernimmen}
\bibinfo{author}{\bibfnamefont{P.}~\bibnamefont{Vernimmen}},
  \bibinfo{author}{\bibfnamefont{P.}~\bibnamefont{Quiry}},
  \bibinfo{author}{\bibfnamefont{M.}~\bibnamefont{Dallocchio}},
  \bibinfo{author}{\bibfnamefont{Y.~L.} \bibnamefont{Fur}}, \bibnamefont{and}
  \bibinfo{author}{\bibfnamefont{A.}~\bibnamefont{Salvi}},
  \emph{\bibinfo{title}{Corporate Finance: Theory and Practice}}
  (\bibinfo{publisher}{John Wiley and Sons, 4th Edition},
  \bibinfo{year}{2014}).

\bibitem[{\citenamefont{Murray}(1994)}]{Murray1994}
\bibinfo{author}{\bibfnamefont{M.~P.} \bibnamefont{Murray}},
  \bibinfo{journal}{The American Statistician}
  \textbf{\bibinfo{volume}{48(1)}}, \bibinfo{pages}{37} (\bibinfo{year}{1994}).

\bibitem[{\citenamefont{N and Larralde}(2016)}]{Mota2016}
\bibinfo{author}{\bibfnamefont{R.~M.} \bibnamefont{N}} \bibnamefont{and}
  \bibinfo{author}{\bibfnamefont{H.}~\bibnamefont{Larralde}},
  \bibinfo{journal}{arXiv:1601.00229}  (\bibinfo{year}{2016}).

\bibitem[{\citenamefont{Delbaen and Schachermayer}(2011)}]{Delbaen2011}
\bibinfo{author}{\bibfnamefont{F.}~\bibnamefont{Delbaen}} \bibnamefont{and}
  \bibinfo{author}{\bibfnamefont{W.}~\bibnamefont{Schachermayer}},
  \bibinfo{journal}{Notices of the AMS} \textbf{\bibinfo{volume}{51(5)}}
  (\bibinfo{year}{2011}).

\bibitem[{\citenamefont{McInnes et~al.}(2018)\citenamefont{McInnes, Healy, and
  Melville}}]{umap}
\bibinfo{author}{\bibfnamefont{L.}~\bibnamefont{McInnes}},
  \bibinfo{author}{\bibfnamefont{J.}~\bibnamefont{Healy}}, \bibnamefont{and}
  \bibinfo{author}{\bibfnamefont{J.}~\bibnamefont{Melville}},
  \bibinfo{journal}{arXiv preprint arXiv:1802.03426}  (\bibinfo{year}{2018}).

\bibitem[{\citenamefont{Goodfellow et~al.}(2014)\citenamefont{Goodfellow,
  Pouget-Abadie, Mirza, Xu, Warde-Farley, Ozair, Courville, and Bengio}}]{GANs}
\bibinfo{author}{\bibfnamefont{I.}~\bibnamefont{Goodfellow}},
  \bibinfo{author}{\bibfnamefont{J.}~\bibnamefont{Pouget-Abadie}},
  \bibinfo{author}{\bibfnamefont{M.}~\bibnamefont{Mirza}},
  \bibinfo{author}{\bibfnamefont{B.}~\bibnamefont{Xu}},
  \bibinfo{author}{\bibfnamefont{D.}~\bibnamefont{Warde-Farley}},
  \bibinfo{author}{\bibfnamefont{S.}~\bibnamefont{Ozair}},
  \bibinfo{author}{\bibfnamefont{A.}~\bibnamefont{Courville}},
  \bibnamefont{and} \bibinfo{author}{\bibfnamefont{Y.}~\bibnamefont{Bengio}},
  in \emph{\bibinfo{booktitle}{Advances in neural information processing
  systems}} (\bibinfo{year}{2014}), pp. \bibinfo{pages}{2672--2680}.

\bibitem[{\citenamefont{Christ et~al.}(2018)\citenamefont{Christ, Braun,
  Neuffer, and Kempa-Liehr}}]{tsfresh}
\bibinfo{author}{\bibfnamefont{M.}~\bibnamefont{Christ}},
  \bibinfo{author}{\bibfnamefont{N.}~\bibnamefont{Braun}},
  \bibinfo{author}{\bibfnamefont{J.}~\bibnamefont{Neuffer}}, \bibnamefont{and}
  \bibinfo{author}{\bibfnamefont{A.~W.} \bibnamefont{Kempa-Liehr}},
  \bibinfo{journal}{Neurocomputing} \textbf{\bibinfo{volume}{307}},
  \bibinfo{pages}{72} (\bibinfo{year}{2018}).

\bibitem[{\citenamefont{Pedregosa et~al.}(2011)\citenamefont{Pedregosa,
  Varoquaux, Gramfort, Michel, Thirion, Grisel, Blondel, Prettenhofer, Weiss,
  Dubourg et~al.}}]{sklearn}
\bibinfo{author}{\bibfnamefont{F.}~\bibnamefont{Pedregosa}},
  \bibinfo{author}{\bibfnamefont{G.}~\bibnamefont{Varoquaux}},
  \bibinfo{author}{\bibfnamefont{A.}~\bibnamefont{Gramfort}},
  \bibinfo{author}{\bibfnamefont{V.}~\bibnamefont{Michel}},
  \bibinfo{author}{\bibfnamefont{B.}~\bibnamefont{Thirion}},
  \bibinfo{author}{\bibfnamefont{O.}~\bibnamefont{Grisel}},
  \bibinfo{author}{\bibfnamefont{M.}~\bibnamefont{Blondel}},
  \bibinfo{author}{\bibfnamefont{P.}~\bibnamefont{Prettenhofer}},
  \bibinfo{author}{\bibfnamefont{R.}~\bibnamefont{Weiss}},
  \bibinfo{author}{\bibfnamefont{V.}~\bibnamefont{Dubourg}},
  \bibnamefont{et~al.}, \bibinfo{journal}{Journal of machine learning research}
  \textbf{\bibinfo{volume}{12}}, \bibinfo{pages}{2825} (\bibinfo{year}{2011}).

\end{thebibliography}
 
\end{document}